\documentclass[11pt]{article}%
\usepackage[authoryear]{natbib}
\usepackage{amsfonts}
\usepackage{amsmath,amssymb,amsthm,enumerate,epsfig,graphicx}
\usepackage{ifthen,latexsym,syntonly}
\usepackage{natbib}
\usepackage{amsmath}
\usepackage{setspace}
\usepackage{amssymb}
\usepackage{graphicx}%
\usepackage{verbatim}
\usepackage{ifthen}
\newtheorem{theorem}{Theorem}

\newtheorem{lemma}{Lemma}

\renewcommand{\cite}{\citet*}

\newcommand{\etal}{{\em et al.}}

\newcommand{\bA}{\mbox{\bf A}}

\newcommand{\bS}{\mbox{\bf S}}

\newcommand{\bX}{\mbox{\bf X}}
\newcommand{\bW}{\mbox{\bf W}}

\newcommand{\bsone}{\mbox{\bf 1}}

\newcommand{\bmu}{\mbox{\boldmath $\mu$}}

\newcommand{\bGamma}{\mbox{\boldmath $\Gamma$}}

\newcommand{\bSigma}{\mbox{\boldmath $\Sigma$}}

\newcommand{\argmin}{\mathrm{argmin}}

\newcommand{\bw}{\mbox{\bf w}}

\begin{document}

\title{  Vast Volatility Matrix Estimation using High Frequency Data for Portfolio Selection
\thanks{The paper was supported by the  NSF Grants DMS-0704337
and  DMS-0714554.  The main part of the work was carried while
Yingying Li was a postdoctoral fellow at Department of Operations
Research and Financial Engineering,
 Princeton University. Yingying Li was further supported by the Research Support Fund from
 the ISOM Department at Hong Kong University of Science and Technology. {\textbf{Address Information:}
Jianqing Fan, Bendheim Center for Finance, Princeton University,
26 Prospect Avenue, Princeton, NJ 08540, USA. 
E-mail: \texttt{jqfan@princeton.edu}. Yingying Li, Department of
 Information Systems, Business Statistics and Operations Management,
 Hong Kong University of Science and Technology, Hong Kong.  E-mail:
\texttt{yyli@ust.hk}. Ke Yu, Department of Operations Research and
Financial Engineering, Princeton University, Princeton, NJ 08540.
E-mail:\texttt{kyu@Princeton.edu}. }}}
\author{Jianqing Fan, Yingying Li, Ke Yu \\ Princeton University, HKUST, Princeton University}
\maketitle





\begin{abstract}
 Portfolio allocation with gross-exposure constraint is an effective
method to increase the efficiency and stability of selected
portfolios among a vast pool of assets, as demonstrated in
\cite{FanZY08}.  The required high-dimensional volatility matrix can
be estimated by using high frequency financial data. This enables us
to better adapt to the local volatilities and local correlations
among vast number of assets and to increase significantly the sample
size for estimating the volatility matrix.  This paper studies the
volatility matrix estimation using high-dimensional high-frequency
data from the perspective of portfolio selection. Specifically, we
propose the use of ``pairwise-refresh time" and ``all-refresh time"
methods proposed by \cite{BHLS08} for estimation of vast covariance
matrix and compare their merits in the portfolio selection.  We also
establish the concentration inequalities of the estimates, which
guarantee desirable properties of the estimated volatility matrix in
vast asset allocation with gross exposure constraints.  Extensive
numerical studies are made via carefully designed simulations.
Comparing with the methods based on low frequency daily data, our
methods can capture the most recent trend of the time varying
volatility and correlation, hence provide more accurate guidance for
the portfolio allocation in the next time period. The advantage of
using high-frequency data is significant in our simulation and
empirical studies, which consist of 50 simulated assets and 30
constituent stocks of Dow Jones Industrial Average index.


\end{abstract}

\section{Introduction}

The mean-variance efficient portfolio theory by \cite{Markowitz52,
Markowitz59} has profound impact on modern finance.  Yet, its
applications to practical portfolio selection face a number of
challenges.  It is well known that the selected portfolios depend
too sensitively on the expected future returns and volatility matrix
\citep{KleinB76, BestG91, ChopraZ93}.  This leads to the puzzle
postulated by \cite{JagannathanM03} why no short-sale portfolio
outperforms the efficient Markowicz portfolio.  See also
\cite{DeroonNW01} on the study of optimal no-short sale portfolio on
emerging market.  The sensitivity on the dependence can be
effectively addressed by the introduction of the constraint on the
gross exposure of portifolios \citep{FanZY08}. In particular,
\cite{FanZY08} shows, with non-asymptotic inequalities, that for a
range of gross exposure constraint parameters, the actual risk of an
empirically selected optimal portfolio, the actual risk of the
theoretically optimal portfolio, and the estimated risk of an
empirically selected optimal portfolio are in fact close.  The
accuracy depends only on the gross exposure parameter and the
maximum componentwise estimation error of expected returns and
covariance matrix --- there is little error accumulation effect. The
results are demonstrated also by both simulation and empirical
studies. This gives not only a theoretical answer to the puzzle
postulated by \cite{JagannathanM03} but also paves a way for optimal
portfolio selection in practice.

The second challenge of the implementation of Markowitz's portfolio
selection theory is the intrinsic difficulty of the estimation of
the large volatility matrix.   This is well documented in the
statistics and econometrics literature even for the static large
covariance matrix
\citep{Johnstone01,BickelL08,FanFL08,LamF09,RothmanLZ09}.  The
additional challenge comes from the time-varying nature of a large
volatility matrix.  For a short and medium  holding period (one day
or one week, say), the expected volatility matrix in the near future
can be very different from the average of the expected volatility
matrix over a long time horizon (the past one year, say).  As a
result, even if we know exactly the realized volatility matrix in
the past, the bias can still be large. This calls for a stable and
robust portfolio selection. The portfolio allocation under the gross
exposure constraint provides a needed solution.  To reduce the bias
of the forecasted expected volatility matrix, we need to shorten the
learning period to better capture the dynamics of the time-varying
volatility matrix, adapting better to the local volatility and
correlation. But this is at the expense of a reduced sample size.
The wide availability of high-frequency data provides sufficient
amount of data for reliable estimation of the volatility matrix.

Recent years have seen dramatic developments in the study of high
frequency data in integrated volatility. Statisticians and
econometricians have been focusing on the interesting and
challenging problem of volatility estimation in the presence of
market microstructure noise and asynchronous tradings, which are the
style features of high-frequency financial data.  The progresses are
very impressive with a large literature. Assuming the price
processes follow Brownian semimartingales to satisfy the
no-arbitrage based characterizations \citep{DelbaenS94}, if there
were no market microstructure noise, and if the processes are
observed synchronously on grids that become denser, classical
results in stochastic calculus show that the realized variance and
realized covariance are consistent estimators of the quadratic
variation and quadratic co-variation of two price processes; see for
example \cite{KaratzasS00} and \cite{JacodS03}. When directly
applied to high-frequency financial data, however, \cite{ABDL00}
show that the realized variance exhibits a large positive bias when
the sampling frequency gets higher, through their famous signature
plots; \cite{Epps79} documented that the correlation estimates based
on the realized covariances tend to be biased toward zero when
sampled at high frequencies. The recent developments have enabled us
to understand much better the signature plots and Epps effect.
Analytical explanations of how the market microstructure noise and
asynchronization may affect the estimates and ways to correct for
the biases have been given. In particular, in the one dimensional
case when the focus is on estimation of integrated volatility,
\cite{AMZ05} discussed a subsampling scheme; \cite{ZhangMA05}
proposed a two-scale estimate which was extended and improved by
\cite{Zhang06} to multiple scales; \cite{FanWang07} separated jumps
from diffusions in presence of market microstructural noise using a
wavelet method; the robustness issues are addressed by \cite{LiM07};
the realized kernel methods are proposed and thoroughly studied in
\cite{BHLS09a, BHLS09b}; \cite{JLMPV09} proposed a pre-averaging
approach to reduce the market microstructral noise; \cite{Xiu08}
demonstrated that a simple quasi-likelihood method achieves the
optimal rate of convergence for estimating integrated volatility.
For estimation of integrated covariation, the non-synchronized
trading issue was first addressed by \cite{HayashiY05} in absence of
the microstructural noise; the kernel method with refresh time idea
was first proposed by \cite{BHLS08}; \cite{Zhang09} extend the
two-scale method to study the integrated covariation using a
previous tick method; \cite{WYZL09} aggregate daily integrated
volatility matrix via a factor model; \cite{AFX} extend the
quasi-maximum likelihood method; \cite{KinnebrockMC09} extend the
pre-averaging technique.

The aim of this paper is to study the volatility matrix estimation
using high-dimensional high-frequency data from the perspective of
financial engineering. Specifically, our main topic is how to
extract the covariation information from high-frequency data for
asset allocation and how effective they are.  Two particular
strategies are proposed for handling the non-synchronized trading:
``pairwise-refresh'' and ``all-refresh'' schemes.  The former
retains much more data points and estimates covariance matrix
componentwise, which is usually not semi-positive definite, whereas
the latter retains far less data points and the resulting covariance
matrix is usually semi-positive definite.  As a result, the former
has a better componentwise estimation error and is better in
controlling risk approximation mentioned in the first paragraph of
the introduction. However, the merits between the two methods are
not that simple.  In implementation, quadratic programming
algorithms require the estimated covariance matrix to be
semi-positive definite.  Therefore, we need to project the estimate
of covariance matrix based on the ``pairwise-refresh'' scheme onto
the space of the semi-positive definite matrices. However, the
projections distort the accuracy of the elementwise estimation.  As
a result, the pairwise-refresh scheme does not have much more
advantage than the all-refresh method, though the former is very
easy to implement. However, both methods significantly outperform
the methods based on low frequency data, since they adapt better to
the time-varying volatilities and correlations.  The comparative
advantage is more dramatic when there are rapid changes of the
volatility matrix over time. This will be demonstrated in both
simulation and empirical studies.

As mentioned in the introduction and demonstrated in
Section~\ref{sec2}, the accuracy of portfolio risk relative to the
theoretically optimal portfolio is governed by the maximum
elementwise estimation error. How does this error grow with the
number of assets? Thanks to the concentration inequalities derived
in this paper, it grows only at the logarithmic order of the number
of assets.  This gives a theoretical endorsement why the portfolio
selection problem is feasible for vast portfolios.

The paper is organized as follows.  Section~\ref{sec2} gives an
overview of portfolio allocation using high-frequency data.
Section~\ref{sec3} studies the volatility matrix estimation using
high-frequency data from the perspective of asset allocation, where
the analytical results are also presented. How well our idea works
in simulation and empirical studies can be found in
Sections~\ref{sec3} and \ref{sec4}, respectively. Conclusions are
given in Section~\ref{sec5}. Technical conditions and proofs are
relegated to the appendix.

\section{Constrained Portfolio Optimization with High Frequency Data}\label{sec2}

\subsection{Problem Setup}

Consider a pool of $p$ assets, with log-price processes $X^{(1)}_t$,
$X^{(2)}_t$, $\cdots$ $X^{(p)}_t$. Denote by $\bX_s =
(X_s^{(1)},\cdots, X_s^{(p)})^T$ the vector of the log-price
processes at time $s$. Suppose they follow an It\^{o} process,
namely,
\begin{equation} \label{b1}
    d \bX_t = \bmu_t dt+\bS_t^{1/2} d\bW_t
\end{equation}
where $\bW_t$ is the vector of $p$-dimensional standard Brownian
motions. The drift vector $\bmu_t$ and the instantaneous variance
$\bS_t$ can be stochastic processes and are assumed to be bounded
and independent of $\bW_t$.

A given portfolio with the allocation vector $\bw$ at time $t$ and a
holding period $\tau$ has the log-return $\bw^T \int_{t}^{t+\tau}
dX_s$ with variance (risk)
\begin{equation}  \label{b2}
    R_{t, \tau}(\bw) = \bw^T \bSigma_{t, \tau} \bw,
\end{equation}
where $\bw^T \bsone = 1$ and
\begin{equation}  \label{b3}
   \bSigma_{t, \tau} = \int_t^{t+\tau} E_t \bS_u du
\end{equation}
with $E_t$ denoting the conditional expectation given the history up
to time $t$. Let $w^+$ be the propotion of long positions and $w^-$
be the proposition of the short positions.  Then, $\|\bw\|_1 = w_+ +
w^-$ is the gross exposure of the portfolio. To simplify the
problem, following \cite{JagannathanM03} and \cite{FanZY08} and
other papers in the literature, we  consider only the risk
optimization problem. In practice, the expected return constraint
can be replaced by the constraints of sectors or industries, to
avoid unreliable estimates of the expected return vector.  For a
short-time horizon, the expected return is usually negligible.
Following \cite{FanZY08}, we consider the following risk
optimization under gross exposure constraints:
\begin{equation}  \label{b4}
    \min \bw^T \bSigma_{t, \tau} \bw, \quad s.t. \| \bw \|_1 \leq c \mbox{ and }
    \bw^T \bsone = 1,
\end{equation}
where $c$ is the total exposure allowed.  Note that using $w^+ - w^-
= 1$, the problem (\ref{b4}) puts equivalently the constraint on the
proportion of the short positions: $w^- \leq (c-1)/2$.

Problem (\ref{b4}) involves the conditional expected volatility
matrix (\ref{b3}) in the future.  Unless we know exactly the dynamic
of the volatility process, this is usually unknown, even if we
observed the entire continuous paths up to the current time $t$.  As
a result, we rely on the approximation even with ideal data that we
were able to observe the processes continuously without error.  The
typical approximation is
\begin{equation} \label{b5}
      \tau^{-1} \bSigma_{t, \tau} \approx h^{-1} \int_{t-h}^{t} \bS_u du,
\end{equation}
for an appropriate window width $h$ and we estimate $\int_{t-h}^{t}
\bS_u du$ based on the historical data at the time interval $[t-h,
t]$.

The approximation (\ref{b5}) holds reasonably well when $\tau$ and
$h$ are both small.  This relies on the continuity assumptions:
local time-varying volatility matrices are continuous in $\tau$. The
approximation is also reasonable when both $\tau$ and $h$ are large.
This relies on the stationarity assumption so that both quantity
will be approximately $E \bS_u$, when the stochastic volatility
matrix $\bS_u$ is stationary.  The approximation is not good when
$\tau$ is small whereas $h$ is large as long as $\bS_u$ is time
varying, whether or not the stochastic volatility $\bS_u$ is
stationary or not.  In other words, when the holding time horizon
$\tau$ is short, as long as $\bS_u$ is time varying, we can only use
a short time window $[t-h, t]$ to estimate $\bSigma_{t, \tau}$.  The
recent arrivals of high-frequency data make this problem feasible.

The approximation error in (\ref{b5}) can not usually be evaluated
unless we have a specific parametric model on the stochastic
volatility matrix $\bS_u$.  However, this is at the risk of model
misspecifications and nonparametric approach is usually preferred
for high-frequency data.   With $p^2$ elements are approximated,
which can be in the order of hundred of thousands or millions, a
natural question to ask is whether these errors accumulate and
whether the result (risk) is stable.  The gross-exposure constraint
gives a stable solution to the problem as shown in \cite{FanZY08}.

\subsection{Risk approximations with gross exposure constraints}

The utility of gross-exposure constraint can easily be seen through
the following inequality. Let $\hat{\bSigma}_{t, \tau}$ be an
estimated covariance matrix and
\begin{equation}
\hat{R}_{t, \tau} (\bw) = \bw^T  \hat{\bSigma}_{t, \tau} \bw
\label{b6}
\end{equation}
be estimated risk of the portfolio. Then, for any portfolio with
gross-exposure $\|\bw\|_1 \leq c$, we have
\begin{eqnarray} \label{b7}
| \hat{R}_{t, \tau} (\bw) - R_{t, \tau} (\bw)| & \leq & \sum_{i=1}^p
\sum_{j=1}^p
           |\hat{\sigma}_{i,j} - \sigma_{i,j}| |w_i| |w_j| \nonumber \nonumber \\
    & \leq & |\bSigma_{t,\tau} - \hat{\bSigma}_{t,\tau}|_\infty \|\bw\|_1^2 \nonumber \\
    & \leq & |\bSigma_{t,\tau} - \hat{\bSigma}_{t,\tau}|_\infty c^2,
\end{eqnarray}
where $\hat{\sigma}_{i,j}$ and $\sigma_{i,j}$ are respectively the
$(i,j)$ elements of $\hat{\bSigma}_{t,\tau}$ and $\bSigma_{t,\tau}$,
and
$$
     |\bSigma_{t,\tau} - \hat{\bSigma}_{t,\tau}|_\infty = \max_{i, j}
     |\hat{\sigma}_{i,j} - \sigma_{i,j}|
$$
is the maximum componentwise estimation error.  The risk
approximation (\ref{b7}) reveals that there is no error accumulation
effect when gross exposure $c$ is moderate.

From now on, we drop the dependence of $t$ and $\tau$ whenever there
is no confusion.  This facilitates the notation.

\cite{FanZY08} showed further that the risks of optimal portfolios
are indeed close. Let
\begin{equation}
 \bw_{opt} = \argmin_{\bw^T \bsone = 1, \; ||\bw||_1 \leq c}
    R(\bw), \qquad  \hat \bw_{opt}= \argmin_{\bw^T \bsone = 1, \; ||\bw||_1 \leq c}
    \hat{R}(\bw)
    \label{b8}
\end{equation}
be respectively the theoretical (oracle) optimal allocation vector
we want and the estimated optimal allocation vector we get.  Then,
$R(\bw_{opt})$ is the theoretical minimum risk and
$R(\hat\bw_{opt})$ is the actual risk of our selected portfolio,
whereas $\hat{R}(\hat\bw_{opt})$ is our perceived risk, which is the
quantity known to financial econometricians. They showed that
\begin{eqnarray}
   |R(\hat\bw_{opt}) - R(\bw_{opt})| & \leq & 2 a_p c^2, \label{b9} \\
   |R(\hat\bw_{opt}) - \hat{R}(\hat{\bw}_{opt})| & \leq & a_p c^2,  \label{b10}\\
   |R(\bw_{opt}) - \hat{R}(\hat \bw_{opt})| &\leq& a_p c^2. \label{b11}
\end{eqnarray}
with $a_p = |\hat{\bSigma} - \bSigma|_\infty$, which usually grows
slowly with the number of assets $p$. These reveal that the three
relevant risks are in fact close as long as the gross-exposure
parameter $c$ is moderate and the maximum componentwise estimation
error $a_p$ is small.

The above risk approximations hold for any estimate of covariance
matrix.  It does not even require $\hat{\bSigma}$ a semi-positive
definite matrix.  This facilitates significantly the method of
covariance matrix estimation.  In particular, the elementwise
estimation methods are allowed.  In fact, since the approximation
errors in (\ref{b9}), (\ref{b10}) and (\ref{b11}) are all controlled
by the maximum elementwise estimation error, it can be advantageous
to use elementwise estimation methods.  This is particularly the
case for the high-frequency data where trading are non-synchronized.
The synchronization can be done pairwisely or for all assets.  The
former retains much more data than the latter, as shown in the next
section.

\section{Estimation of Covariance Matrix Using High Frequency Data} \label{sec3}

\subsection{All-refresh method and pairwise-refresh method}

Estimating high-dimensional volatility matrix using high-frequency
data is a challenging task. One of the challenges is the
non-synchronicity of trading. Several synchronization schemes have
been proposed. The refresh time method is proposed in \cite{BHLS08}
and the previous tick method is proposed in \cite{Zhang09}. The
former uses more efficiently the available data and will be used in
this paper.

The idea of refresh time is to wait until all assets are traded at
least once at time $v_1$ (say) and then use the last price traded
before or at $v_1$ of each asset as its price at time $v_1$. This
obtains one synchronized price vector at time $v_1$.  The clock now
starts again.  Wait until all assets are traded at least once at
time $v_2$ (say) and again use the previous tick price of each asset
as its price at time $v_2$. This yields the second synchronized
price vector at time $v_2$. Repeat the process until all available
trading data are synchronized. Clearly, the process discards a large
portion of the available trades:  After each synchronization, we
always wait until the slowest stock to trade once.  But this is the
most efficient synchronization scheme. We will refer this
synchorization scheme as the ``all-refresh time'' (The method is
called all-refresh method for short). \cite{BHLS08} advocate the
kernel method to estimate integrated volatility matrix after
synchronization, but this can also be done by using other methods.
The advantage of the all-refresh method is that the estimated
covariance matrix can be made semi-positive definite.

A more efficient method to use the available sample is the pairwise
refresh time scheme, which synchronizes the trading for each pair of
assets separately (The method is called pairwise-refresh method for
short). This retains far more data points, but we have to estimate
the covariance matrix elementwise.  The resulting covariance matrix
is not necessarily semi-positive definite.  Thanks to the gross
exposure constraint, this is still applicable to the portfolio
selection problems, as long as the elementwise estimation error is
small. See (\ref{b7}) -- (\ref{b11}).  The pairwise-refresh scheme
makes far more efficient use of the rich information in
high-frequency data, and enables us to estimate each element in the
volatility matrix more precisely, which helps improve the efficiency
of the selected portfolio. We will study the merits of these two
methods.

The pairwise estimation method allows us to use a wealth of
univariate integrated volatility estimators, such as the  two-scale
realized volatility (TSRV) \citep{ZhangMA05}, the multi-scale
realized volatility (MSRV) \citep{Zhang06}, the wavelet method
\citep{FanWang07}, the realized kernel method \citep{BHLS09a,
BHLS09b}, the pre-averaging approach \citep{JLMPV09} and the QMLE
method \citep{Xiu08}.  For any given two assets with log-price
processes $X^{(i)}_t$ and $X^{(j)}_t$, with pairwise-refresh times,
the synchronized prices of $X^{(i)}_t + X^{(j)}_t$ and $X^{(i)}_t -
X^{(j)}_t$ can be computed.  With the univariate estimate of the
integrated volatilities $<X^{(i)} + X^{(j)}>$ and $<X^{(i)} -
X^{(j)}>$, the integrated covariantion can be estimated as
\begin{equation} \label{b12}
\hat{\sigma}_{i,j} = \langle X^{(i)}, Y^{(j)} \rangle = ( \langle
X^{(i)} + X^{(j)} \rangle - \langle X^{(i)} - X^{(j)} \rangle )/4.
\end{equation}
In particular, the diagonal elements are estimated by the method
itself. When the TSRV is used, this results in the two-scale
realized covariance (TSCV) estimate \citep{Zhang09}.

\subsection{Pairwise refresh method and TSCV\label{sec3.2}}

We now focus on the pairwise estimation method.
 To
facilitate the notation, we reintroduce it.

We consider two log price processes $X$ and $Y$ that satisfy
\begin{equation} \label{b13}
  dX_t = \mu_t^{(X)} dt + \sigma_t^{(X)} dB^{(X)}
  \qquad \mbox{ and } \quad dY_t = \mu_t^{(Y)} dt + \sigma_t^{(Y)} dB^{(Y)},
\end{equation}
where $cor(B^{(X)}_t, B^{(Y)}_t)=\rho_t^{(X,Y)}$. For the two
processes $X$ and $Y$, consider the problem of estimating $\langle
X, Y \rangle_T$ with $T=1$. Denote by $\mathcal{T}_n$ the
observation times of $X$ and $\mathcal{S}_m$ the observation times
of $Y$. Denote the elements in $\mathcal{T}_n$ and $\mathcal{S}_m$
by $\{\tau_{n,i}\}_{i=0}^n$ and $\{\theta_{m,i}\}_{i=0}^m$
respectively, in an ascending order ($\tau_{n,0}$ and $\theta_{m,0}$
are set to be 0). We assume that the actual log-prices are not
observable, but are observed with microstructure noises:
\begin{equation}\label{b14}
X^o_{\tau_{n,i}}   = X_{\tau_{n,i}}+\epsilon_{i}^X, \quad \mbox{and}
\quad Y^o_{\theta_{m,i}} = Y_{\theta_{m,i}}+\epsilon_{i}^Y
\end{equation}
where $X^o$ and $Y^o$ are the observed transaction prices in the
logarithmic scale, and  $X$ and $Y$ are the latent log prices govern
by the stochastic dynamics (\ref{b13}). We assume that the
microstructure noise $\epsilon_i^X$ and $\epsilon_i^Y$ processes are
independent of the $X$ and $Y$ processes and that
\begin{equation}\label{b15}
    \epsilon_i^{X}\sim_{i.i.d.}
    N(0,\eta_X^2)\;\; \mbox{ and } \;\; \epsilon_i^{Y}\sim_{i.i.d.}
    N(0,\eta_Y^2).
\end{equation}
Note that this assumption is mainly for the simplicity of
presentation; as we can see from the proof, one can for example
easily replace the Gaussian assumption with the sub-Gaussian
assumption without affecting our results.

The pairwise refresh time
$\mathcal{V}=\{v_0,v_1,\cdots,v_{\tilde{n}}\}$ can be obtained by
setting $v_0=0$, and
$$
  v_i=\max\big\{\min\{\tau\in\mathcal{T}_n:\tau>v_{i-1}\},\; \min\{\theta\in\mathcal{S}_m:\theta>v_{i-1}\}\big\},
$$
where  $\tilde{n}$ is the total number of refresh times in the
interval $(0,1]$. The actual sample times for the two individual
processes $X$ and $Y$ that correspond to the refresh times are
$$
t_i=\max\{\tau\in\mathcal{T}_n:\;\tau\leq v_i\}\;\mbox{ and }
s_i=\max\{\theta\in\mathcal{S}_m:\;\theta\leq v_i\},
$$
which is really the previous-tick measurement.

We study the property of the TSCV based on the {\it asynchronous}
data:
\begin{equation} \label{b16}
    \widehat{\langle X,Y \rangle}_1=[X^o,
    Y^o]_1^{(K)}-\frac{\bar{n}_K}{\bar{n}_J}[X^o,
    Y^o]_1^{(J)},
\end{equation}
where $$[X^o, Y^o]_1^{(K)}=\frac{1}{K}\sum_{i=K}^{\tilde
n}(X_{t_i}^o-X_{t_{i-K}}^o)(Y_{s_i}^o-Y_{s_{i-K}}^o)$$ and
$\bar{n}_K=(\tilde n-K+1)/K$. As discussed in \cite{Zhang09}, the
optimal choice of $K$ has order $K=O({\tilde n}^{2/3})$, $J$ can be
taken to be a constant such as $1$. In the following analysis, we
consider the specific case when
$$J=1 \;(\mbox{or } \bar{n}_J={\tilde n})\;\; \mbox{ and } \;\;\bar{n}_K=O({\tilde n}^{1/3}).$$

When either the microstructure error or the asynchronicity exists,
the realized covariance is seriously biased. An asymptotic normality
result in \cite{Zhang09} reveals that TSCV can simultaneously remove
the bias due to the microstructure error and the bias due to the
asynchronicity. However, this result is not adequate for our
application to the vast volatility matrix estimation. The maximum
componentwise estimation error $a_p$ depends on the number of assets
$p$.  To understand its impact on $a_p$, we need to establish the
concentration inequality. In particular,  for a sufficiently large
$|x| = O((\log p)^a)$, if
\begin{equation} \label{b17}
   \max_{i, j} P\{ \sqrt{n} | \sigma_{ij} - \hat{\sigma}_{ij}| > x \} <
      \exp( - C x^{1/a}),
\end{equation}
for two positive constants $a$  and $C$, then
\begin{equation}
a_p = |\bSigma - \hat{\bSigma} |_\infty = O_P \left ( \frac{ (\log
p)^{a}}{\sqrt{n}} \right ).   \label{b18}
\end{equation}
We will show  in the next section that the result indeed holds with
$a = 1/2$ and $n$ replaced by the minimum subsample size. Hence the
impact of the number of assets is limited, only of the logarithmic
order.

\subsection{Concentration Inequalities \label{sec3.3}}

Inequality (\ref{b17}) requires the conditions on both diagonal
elements and off-diagonal elements.  Technically, they are treated
differently. For the diagonal cases, the problem corresponds to the
estimation of integrated volatility and there is no issue of
asynchronicity.  TSCV (\ref{b16}) reduces to TSRV \citep{ZhangMA05},
which is explicitly given by

\begin{equation} \label{b19}
    \widehat{\langle X,X \rangle}_1=[X^o, X^o]_1^{(K)}
    -\frac{\bar{n}_K}{\bar{n}_J}[X^o, X^o]_1^{(J)},
\end{equation}
where
$$
    [X^o, X^o]_1^{(K)}=\frac{1}{K}\sum_{i=K}^{n}(X_{t_i}^o-X_{t_{i-K}}^o)^2
$$
and $\bar{n}_K=(n-K+1)/K.$

As shown in \cite{ZhangMA05}, the optimal choice of $K$ has order
$K=O(n^{2/3})$ and $J$ can be taken to be a constant such as $1$.
Again, for the TSRV, in the following analysis, we will consider the
specific case when $J=1$ (or $\bar{n}_J=n$) and
$\bar{n}_K=O(n^{1/3})$.

To facilitate the reading, we relegate the technical conditions and
proofs to the appendix. The following two results establish the
concentration inequalities for the integrated volatility and
integrated covariation.

{\theorem  \label{thm1} Let $X$ process be as in (\ref{b13}), and
$n$ be the total number of observations for the $X$  process  during
the time interval (0,1]. Under Conditions 1-4 in section
\ref{secA.1}, for $x\in[0, c {n}^{1/6}]$,
$$
  P \Big \{n^{1/6}|\widehat{\langle X,X \rangle}_1-\int_0^1{\sigma_t^{(X)}}^2 dt|>x
  \Big \} \leq  4 \exp \{-C x ^{2}\}$$ for positive constants $c$ and $C$. A
set of candidate values for $c$ and $C$ are given in (\ref{b49}) for
the case when the TSRV parameters are chosen according to Condition
5.}


{\theorem \label{thm2} Let $X$ and $Y$ processes be as in
(\ref{b13}), and $\tilde n$ be the total number of refresh times for
the processes $X$ and $Y $ during time interval (0,1]. Under
Conditions 1-5 in section \ref{secA.1}, for $x\in[0, c
{\tilde{n}}^{1/6}]$,
$$P\{{\tilde{n}}^{1/6}|\widehat{\langle X,Y \rangle}_1-\int_0^1\sigma_t^{(X)}\sigma_t^{(Y)}\rho_t^{(X,Y)}dt|>x\}
\leq 8 \exp \{-C x ^{2}\}$$ for positive constants $c$ and $C$. A
set of candidate values for $c$ and $C$ are given in (\ref{MC_tscv})
for the case when the TSCV parameters are chosen according to
Condition 5.}

\subsection{Error rates on risk approximations}

Having had the above concentration inequalities, we can now readily
give an upper bound of the risk approximations. Consider the $p$
log-price processes as in Section 2.1. Suppose the processes are
observed with the market microstructure noises. Let $\tilde
n^{(i,j)}$ be the observation frequency obtained by the
pairwise-refresh method for two processes $X^{(i)}$ and $X^{(j)}$
and $\tilde n_*$ be the observation frequency obtained by the
all-refresh method.  Clearly, $\tilde n^{(i,j)}$ is typically much
larger than $\tilde n_*$. Hence, most elements are estimated more
accurately using the pairwise-refresh method than using the
all-refresh method. On the other hand, for less liquidly traded
pairs, its observation frequency of pairwise-refresh time can not be
an order of magnitude larger than $\tilde n_*$.

Using (\ref{b18}), an application to Theorems 1 and 2 to each
element in the estimated integrated covariance matrix yields
\begin{equation}  \label{b20}
    a_p^{\mbox{\scriptsize pairwise-refresh}} = |\hat{\bSigma}^{\mbox{\scriptsize pairwise}} - \bSigma|_\infty = O_P \left ( \frac{\sqrt{\log p}}{\tilde{n}_{\min}^{1/6}} \right ),
\end{equation}
where $\tilde{n}_{\min} = \min_{i, j} {\tilde n}^{(i,j)}$ be the
minimum number of observations of the pairwise-refresh time.

Note that based on our proofs which don't rely on particular
properties of pairwise-refresh times, our results of Theorem 1 and
Theorem 2 are applicable to all-refresh method as well, with the
observation frequency of the pairwise-refresh times replaced by that
of the all-refresh times. Hence,  using the all-refresh time scheme,
we have
\begin{equation} \label{b21}
  a_p^{\mbox{\scriptsize all-refresh}} = |\hat{\bSigma}^{\mbox{\scriptsize all-refresh}}
  - \bSigma|_\infty = O_P \left ( \frac{\sqrt{\log p}}{\tilde n_{*}^{1/6}} \right ).
\end{equation}
Clearly, $\tilde n_{\min}$ is larger than $\tilde n_{*}$.   See
Figure~\ref{Fig2}. Hence, the pairwise refresh method gives a
somewhat more accurate estimate in terms of the maximum elementwise
estimation error.

\subsection{Projections of estimated volatility matrices}

The risk approximations (\ref{b9})-(\ref{b11}) hold for any
solutions to (\ref{b8}) whether the matrix $\hat{\bSigma}$ is
positive semi-definite or not.  However,  convex optimization
algorithms typically require the positive semi-definiteness of the
matrix $\hat{\bSigma}$. Yet, the estimates based on the elementwise
estimation sometimes can not satisfy this and even the one from
all-refresh method can have the same problem if TSRV is used. This
leads to the issue of how to project a symmetric matrix onto the
space of positive semi-definite matrices.

There are two intuitive methods for projecting a $p \times p$
symmetric matrix $\bA$ onto the space of positive semi-definite
matrices. Consider the singular value decomposition:  $\bA =
\bGamma^T \mbox{diag}(\lambda_1, \cdots, \lambda_p) \bGamma$, where
$\bGamma$ is an orthogonal matrix, consisting of $p$ eigenvectors.
The two intuitive appealing projection methods are
\begin{equation} \label{b22}
    \bA_1^+ = \bGamma^T \mbox{diag}(\lambda_1^+, \cdots, \lambda_n^+) \bGamma,
\end{equation}
where $\lambda_j^+$ is the positive part of $\lambda_j$ and
\begin{equation}\label{b23}
    \bA_2^+ = (\bA + \lambda_{\min}^- I_p) / (1 + \lambda_{\min}^-),
\end{equation}
where $\lambda_{\min}^-$ is the negative part of the minimum
eigenvalue.  For both projection methods, the eigenvectors remain
the same as those of $\bA$.  When $\bA$ is positive semi-definite
matrix, we have obviously that $\bA_1 = \bA_2 = \bA$.

In applications, we apply the above transformations to the estimated
correlation matrix $\bA$ rather than directly to the volatility
matrix estimate $\hat \bSigma$.  The correlation matrix $\bA$ has
diagonal elements of 1. The resulting matrix under the projection
method (\ref{b23}) apparently still satisfies this property, whereas
the one under the projection method (\ref{b22}) does not. As a
result, the projection method (\ref{b23}) keeps the integrated
volatility of each asset intact.

In the simulation and empirical studies, we applied both
projections. It turns out that there is no significant difference
between the two projection methods in terms of result. We decided to
apply only the projection (\ref{b23}) in all numerical studies, as
it keeps the individual volatility estimate intact.

\subsection{Comparisons between pairwise-refresh and all-refresh methods}

The pairwise-refresh method keeps far richer information in the
high-frequency data than the all-refresh method. See
Figure~\ref{Fig2}. Thus, it is expected to estimate each element
more precisely. Yet, the estimated correlation matrix is typically
not positive semi-definite. As a result, projection (\ref{b23}) can
distort the accuracy of elementwise estimation.  On the other hand,
the all-refresh method is typically positive semi-definite or nearly
so. The property (\ref{b12}) typically entails the positive
semi-definiteness property, as long as the volatility estimator for
$\langle X, X \rangle$ is always nonnegative. For example, using the
realized kernel method as the building block, the positive
semi-definite version can easily be obtained. Therefore, the
projection (\ref{b23}) has less impact on the all-refresh method
than on the pairwise-refresh method.

Risk approximations (\ref{b9})--(\ref{b11}) are only the upper
bounds. The upper bounds are controlled by $a_p$, which has rates of
convergence govern by (\ref{b20}) and (\ref{b21}). While the average
number of observations of pairwise-refresh time is far larger than
the number of observations $\tilde{n}_{*}$ of the all-refresh time,
the minimum number of observations of pairwise-refresh time
$\tilde{n}_{\min}$ is not much larger than $\tilde{n}_{*}$.
Therefore, the upper bounds (\ref{b20}) and (\ref{b21}) are
approximately of the same order. This together with the distortion
due to projection do not leave much advantage for the
pairwise-refresh method.

\section{Simulation Studies}  \label{sec4}

\newboolean{flagFourFig}
\setboolean{flagFourFig}{true} \setboolean{flagFourFig}{false}

In this section, we simulate the market trading data using a
reasonable stochastic model. As the latent prices and dynamics of
simulations are known, our study on the risk profile is facilitated.
It is a good tool to verify our theoretical results and to quantify
the finite sample behaviors. In particular, we would like to
demonstrate that high frequency data based approaches have a better
risk profile than those based on the low frequency data.

Throughout this paper, the risk is referring to the standard
deviation of portfolio's returns. To avoid ambiguity, we call
$\sqrt{R(\bw_{opt})}$ the theoretical optimal risk or oracle risk,
$\sqrt{R_n(\hat{\bw}_{opt})}$ the perceived optimal risk, and
$\sqrt{R(\hat{\bw}_{opt})}$ the actual risk of the perceived optimal
allocation. 

\subsection{Design of Simulations}

A slightly modified version of the simulation model in \cite{BHLS08}
is used to generate the latent price processes of $p$ traded assets.
It is a multivariate factor model with stochastic volatilities.
Specifically, the latent log-prices $X_t^{(i)}$ follow
\begin{equation} \label{b24}
    dX_t^{(i)} = \mu^{(i)}dt + \rho^{(i)} \sigma_t^{(i)} dB_t^{(i)} + \sqrt{1-(\rho^{(i)})^2} \sigma_t^{(i)} dW_t + \lambda^{(i)} dZ_t,
    \quad i = 1, \cdots, p,
\end{equation}
where the elements of $B$, $W$ and $Z$ are independent standard
Brownian motions. The spot volatility obeys the independent
Ornstein-Uhlenbeck processes:
\begin{equation} \label{b25}
    d\varrho_t^{(i)} = \alpha^{(i)} (\beta_0^{(i)} - \varrho_t^{(i)}) dt + \beta_1^{(i)} dU_t^{(i)},
\end{equation}
where $\varrho_t^{(i)} = \log \sigma_t^{(i)} $ and $U_t^{(i)}$ is an
independent Brownian motion.  The stationary distribution is given
by $N\Bigl(\beta_0^{(i)}, [\beta_1^{(i)}]^2/(2 \alpha^{(i)}  )\Bigr
) $. The integrated quadratic variation and covariation are given by
\begin{eqnarray*}
  \langle X^{(i)} \rangle _t
 & = &  \int_{0}^{t} (\sigma_s^{(i)})^2 ds + \lambda^{(i)} t, \\
 \langle X^{(i)},X^{(j)} \rangle_t & = & \int_{0}^{t} \sqrt{1-(\rho_s^{(i)})^2}
\sqrt{1-(\rho_s^{(j)})^2} \sigma_s^{(i)} \sigma_s^{(j)} ds.
\end{eqnarray*}
The analytic formula for the conditional covariance matrix
$\bSigma_{t, \tau}$ in (\ref{b3}) can be found for our model, but we
decide not to report it for brevity.

%
%

The number of assets $p$ is taken to be 50. Slightly modified from
\cite{BHLS08}, the parameters is set to be $(\mu^{(i)},
\beta_0^{(i)}, \beta_1^{(i)}, \alpha^{(i)}, \rho^{(i)}) =
(0.03x_1^{(i)},$ $-x_2^{(i)},$ $0.75x_3^{(i)},$ $-1/40x_4^{(i)},
-0.7)$ where $x_j^{(i)}$ is an independent realization form the
uniform distribution on  $[0.7,1.3]$. The parameters are kept fixed
during the simulations. In addition, $\lambda^{(i)} =
\exp(\beta_0^{(i)})$, which makes the volatility matrix well
conditioned.

\begin{figure}[t]   
\begin{center}
\begin{tabular}{c c}
\includegraphics[scale=0.6]{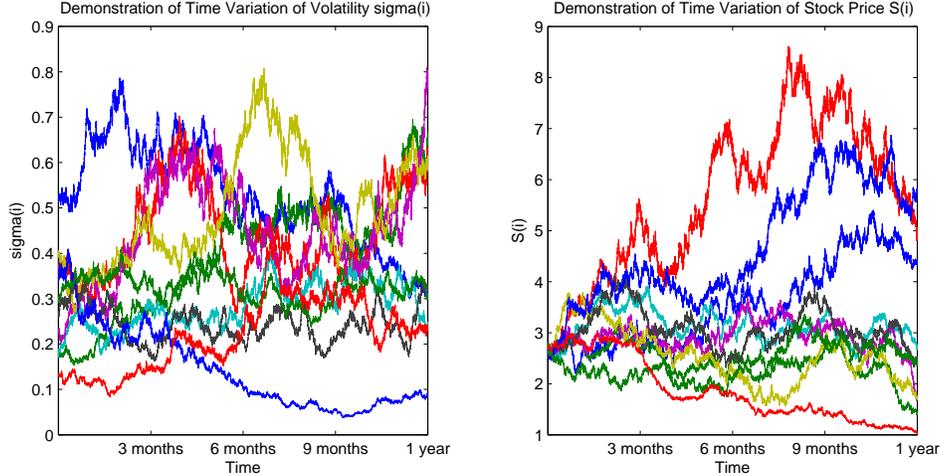} &
\end{tabular}
   \caption{The volatility and asset price processes of $10$ simulated assets.}
\label{Fig1}
\end{center}
\end{figure}

The model (\ref{b24}) is used to generate the latent log-price
values with initial values $X_0^{(i)} = 1$ (log-price) and
$\varrho_0^{(i)}$ from its stationary distribution. The Euler scheme
is used to generate latent price at the frequency of once per
second. To account for the market microstructure noise,  the
Gaussian noises $\varepsilon_t^{(i)} \sim N(0,\omega^2)$ with
$\omega = 0.0005$ are added. Therefore, like (\ref{b14}), the
observed log-prices are $X_t^{o(i)} = X_t^{(i)} +
\varepsilon_t^{(i)}$.  To gain a sense of the extent to which the
asset volatilities $\sigma_t^{(i)}$ and prices $P_t^{(i)}$
($=\exp(X_t^{o(i)})$) vary through time, we plot demonstrative
graphs of $10$ assets' volatility and price processes over a year in
Figure \ref{Fig1}.

To model the non-synchronicity, $p$ independent Poisson processes
with intensitive parameters $\lambda_1,\lambda_2,\cdots,\lambda_p$
are used to simulate the trading times of the assets. Motivated by
the US equity trading dataset (the total number of seconds in a
common trading day of the US equity is $23400$), we set the trading
intensity parameters $\lambda_i$'s to be $0.02i \times 23400$ for $i
= 1,2,\cdots,50$, meaning that the average numbers of trading times
for each asset are spread out in the arithmetic sequence of the
interval $[468, 23400]$.

\subsection{An oracle investment strategy and risk assessment}

An oracle investment strategy is usually a decent benchmark for
other portfolio strategies to be compared with.  There are several
oracle strategies. The one we choose is to make portfolio allocation
based on the covariance matrix estimated using latent prices at the
finest grid (one per second). Latent prices are the noise-free
prices of each asset at every time points (one per second), which
are unobservable in practice and is available to us only in the
simulation.  Therefore, for each asset, there are $23400$ latent
prices in a normal trading day. We will refer to the investment
strategy based on the latent prices as the oracle or latent
strategy. This strategy is not available for the empirical studies.

The assessment of risk is based on the high-frequency data.  For a
given portfolio strategy, its risk is computed based on the latent
prices at the finest grid (one per second) for the in-the-sample
simulation studies; its risk is computed based on the latent prices
at every 15 minutes for the out-of-sample simulation studies;
whereas for the empirical studies, the observed prices at every 15
minutes are used to assess its risk. This mitigates the influence of
the microstructure noises. For the empirical study, we do not hold
positions overnight therefore are immune to the overnight price
jumps (we will discuss the details in Section~\ref{sec5}).

\subsection{In-sample Risk Approximation and Optimal Allocation}

Based on the past $h = 1$ day, the latent prices (at the finest
grid) based estimated TSCV covariance matrix (called latent
covariance for short) is regarded as the true covariance matrix.
There are several methods for estimating covariance matrix based on
observed non-synchronized high-frequency data with microstructure
noise.  In particular, we employ all-refresh method based TSCV
covariance matrix (called all-refresh TSCV covariance), all-refresh
method based realized kernel covariance matrix (called all-refresh
RK covariance, for short), and pairwise-refresh method based TSCV
covariance matrix (called pairwise-refresh TSCV covariance). The
all-refresh RK covariance is included since it is positive
semi-definite and there is no distortion effect due to projection.
The latent covariance serves as the oracle covariance matrix from
which the actual portfolio risk of any portfolio is computed. The
conditioning number of the latent covariance of the $p = 50$ assets
ranges from $192.27$ to $226.46$, with median $210.34$, across $
100$ simulations. The medians of the minimum and maximum eigenvalues
are respectively $0.0004$ and $0.0838$. For the all-refresh RK
approach, the bandwidth of the realized kernel $H$ is
chosen to be $1$, which gives the best risk profile in our numerical analysis. \\

\begin{figure}[t]   
\begin{center}
\begin{tabular}{c c}
\includegraphics[scale=0.7]{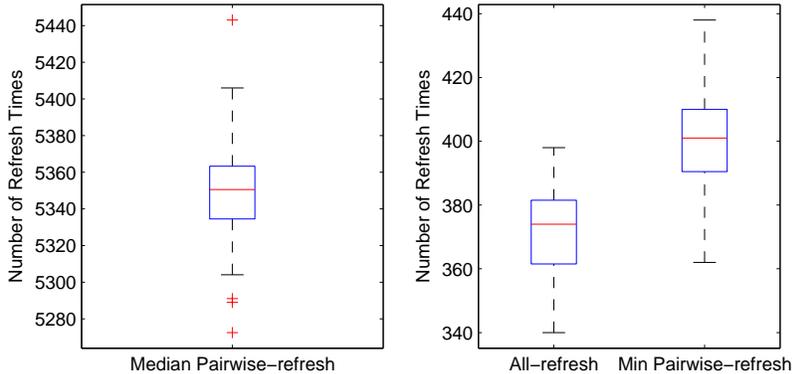} &
\end{tabular}
   \caption{The distributions (from left to right) of the median number of pairwise-refresh times, the minimum number of pairwise-refresh times, and the number of all-refresh times per day across $100$ simulations.}
\label{Fig2}
\end{center}
\end{figure}

The efficiencies of using the rich high-frequency data between
pairwise-refresh and all-refresh methods are contrasted.  In
particular, for each realization, we compute the median number of
pairwise-refresh times $\mbox{median}_{i,j}(\tilde{n}^{(i,j)})$, the
minimum number of pairwise-refresh times $M_{\min} =
\min_{i,j}(\tilde{n}^{(i,j)})$ (see (\ref{b20})), and the number of
all-refresh times $\tilde{n}_*$ (see (\ref{b21})).  The
distributions of these three numbers are summarized in
Figure~\ref{Fig2}.  It is clear that the pairwise refresh scheme
uses far more data on average, yet minimum number of
pairwise-refresh time is not appreciably larger than that of refresh
time.

To gain insights on the risk approximations, we consider 4 specific
portfolios of the $p = 50$ assets with the weight vectors
\begin{eqnarray*}
\begin{array}{l l l}
 w_1   = (\frac{1}{p},\frac{1}{p},\cdots,\frac{1}{p})^T,& & w_2 =
(1,0,\cdots,0)^T,\\
 w_3   =
(\frac{b}{2}-\frac{1}{2}+\frac{2}{p},\frac{1}{2}-\frac{b}{2},\frac{1}{p},\cdots,\frac{1}{p})^T,
 &  & w_4 = (\frac{1}{2}+\frac{b}{2},\frac{1}{2}-\frac{b}{2},0,\cdots,0)^T
\end{array}
\end{eqnarray*}
with $b = 3$.  Their daily risks are computed based on various
covariance estimators and are compared with the actual risk, which
is computed based on the latent price.  This is done across 100
simulations.  The medians, robust standard deviation (defined as
interquartile range divided by 1.35) and other characteristics are
summarized in Table \ref{tab1}.

\begin{table}[h]             
\begin{center}
\caption{\bf Risk approximation for $p = 50$ and $n = 100$}
\label{tab1}
\end{center}
\vspace*{-0.3 in}

\noindent\small We used the high frequency data for $100$
independent trading days. The covariance of the $50$ stocks is
estimated according to various estimators. These estimated
covariance matrices are used to compute the perceived risks of $4$
portfolios. Relevant statistics are recorded. (All the
characteristics are annualized.)
\begin{center}
\small \begin{tabular} {lcccc} \hline

\multicolumn{5}{c}{\em { Median and Robust Standard Deviation (RSD) of Risk}} \\
\hline
& Latent & All-refresh TSRV & All-refresh RK & Pairwise TSRV \\
Portfolio & Median(RSD) & Median(RSD) & Median(RSD) & Median(RSD)
 \\ \hline
 $w_1$                  &  0.4408 (0.0032) & 0.3875 (0.1075) & 0.4343 (0.0241) & 0.4192 (0.0690)\\
 $w_2$                  &  0.5916 (0.0060) & 0.5229 (0.1259) & 0.6230 (0.0258) & 0.5936 (0.1285)\\
 $w_3$                  &  0.5399 (0.0044) & 0.4694 (0.0907) & 0.5833 (0.0255) & 0.5202 (0.0736)\\
 $w_4$                  &  0.8442 (0.0077) & 0.7531 (0.1748) & 0.9228 (0.0418) & 0.8390 (0.1789)\\
 \hline
\multicolumn{5}{c}{\em {Median and RSD of Absolute Risk Difference from the Oracle (Latent)}} \\
\hline
& All-refresh TSRV & All-refresh RK & Pairwise TSRV \\
Portfolio & Median(RSD) & Median(RSD) & Median(RSD)
 \\ \hline
 $w_1$                  &  0.0889 (0.0769) & 0.0183 (0.0153) & 0.0547 (0.0439)\\
 $w_2$                  &  0.1054 (0.0700) & 0.0344 (0.0272) & 0.0804 (0.0813)\\
 $w_3$                  &  0.0936 (0.0665) & 0.0437 (0.0300) & 0.0599 (0.0593)\\
 $w_4$                  &  0.1470 (0.1022) & 0.0794 (0.0393) & 0.1089 (0.0941)\\
 \hline
 \multicolumn{5}{c}{\em {Median and RSD of $L_{1}$ Norm of Absolute Covariance Difference ($a_p$)}} \\
\hline
& All-refresh TSRV & All-refresh RK & Pairwise TSRV \\
Portfolio & Median(RSD) & Median(RSD) & Median(RSD)
 \\ \hline
                        &  0.2476 (0.1460) & 0.0603 (0.0270) & 0.1730 (0.0746)\\
 \hline
\end{tabular}
\end{center}
\end{table}

From the result, we can see that both the all-refresh TSRV and
pairwise-refresh TSRV methods, especially all-refresh TSRV method,
have a tendency to underestimate the risk in comparison with the
latent risks, while all-refresh RK method has a tendency to
overestimate the risk. In terms of the absolute risk difference from
the oracle, for 3 out of the 4 portfolios, pairwise-refresh TSRV
method outperforms the all-refresh TSRV method. The same
relationship can be observed when we turn to the $L_{1}$ norm of the
absolute covariance difference ($a_p$) as well.  The RK method
outperforms the TSRV method. These are in line with our expectation.

\begin{figure}[t]   
\begin{center}
\begin{tabular}{c c}
\includegraphics[scale=0.7]{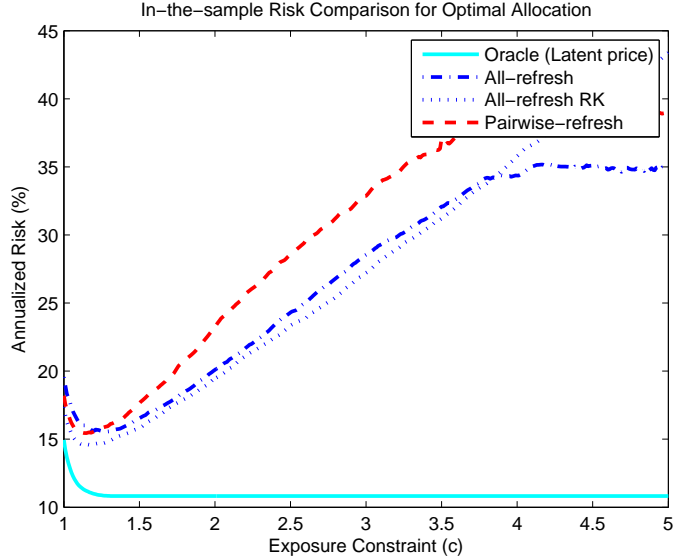} &
\end{tabular}
   \caption{The medians of the actual risks of the in-sample optimal allocations based on the high-frequency estimated covariance matrices using 1 trading day's intra-day data($p = 50$, $n = 100$).}
   \label{Fig3}
\end{center}
\end{figure}

We now study the problem of the optimal portfolio allocation under
gross exposure constraints.  The optimal allocation vectors are
computed based on the latent covariance, all-refresh TSCV
covariance, all-refresh RK covariance and pairwise-refresh TSCV
covariance and their actual risks are computed based on the latent
covariance matrix. The medians of these actual risks against the
gross exposure parameter $c$ are depicted in Figure~\ref{Fig3}.

Firstly, the all-refresh RK method outperforms the two TSRV methods
when the gross exposure is below $3.7$. The pairwise-refresh TSRV
method outperforms the all-refresh TSRV method where the gross
exposure is smaller than $1.2$. That agrees with what we expected
since the smaller the gross exposure is, the tighter the bound
(\ref{b9}) on the risk difference. It is obvious that the
pairwise-refresh TSRV method gives an estimated covariance matrix
with higher element-wise accuracy than the all-refresh TSRV method,
therefore the former outperforms the latter where the bound is the
tightest (gross exposure below $1.2$) for this simulation design.

Secondly, all the methods produce an upward-sloping risk curve up to
some point and an almost flat curve beyond that (the curve is
clipped). This is mainly due to the fact that we use only the
intra-day data for 1 trading day, which does not result in
sufficient amount of data to yield a stable estimate of the $50
\times 50 $ covariance matrix. As the result, the estimated
covariance matrix can be ill-conditioned. As $c$ increases, the
selected portfolios become increasingly unstable. When $c$ reaches 5
or so, the selected portfolio becomes basically a randomly selected
portfolio.  Hence, their actual risks become larger and flat
afterwards.

\subsection{Out-of-sample Optimal Allocation}

One of the main purposes of this paper is to investigate the
comparative advantage of the high frequency based methods against
the low frequency based method (especially in the context of
portfolio investment). Hence, it is essential for us to run the
following out-of-sample investment strategy test which includes both
the high frequency and low frequency based approaches. Moreover,
since in the empirical studies, we do not know the latent asset
prices, the out-of-sample test should be designed so that it can
also be conducted  in the empirical studies.

We simulate the prices of $50$ traded assets using the model
(\ref{b24}) and (\ref{b25}) with microstructure noise for the
duration of $200$ trading days (numbered as day $1$, day $2$, ...,
day $200$) and record all the tick-by-tick trading times and trading
prices of the assets. We assume that there are no overnight jumps
for asset prices, meaning one trading day's closing price of an
asset is always the same as the next trading day's opening price of
that asset.

We start investing $1$ unit of capital into the pool of assets with
low frequency and high frequency based strategies from day $101$
(the portfolios are bought at the opening of day $101$). For the low
frequency strategy, we use the previous $100$ trading days' daily
closing prices to compute the sample covariance matrix and make the
portfolio allocation accordingly with the gross exposure
constraints. For the all-refresh high frequency strategies, we use
the previous $h = 10$ trading days' tick-by-tick trading data, use
all-refresh time to synchronize the trades of the assets before
applying realized kernel and TSCV to estimate the integrated
volatility matrix and make the portfolio allocation, while for the
pairwise-refresh high frequency strategy, we use pairwise-refresh
times to synchronize each pair of assets and apply TSCV to estimate
the integrated covariance for the corresponding pair. With the
projection technique (\ref{b23}), the resulting TSCV integrated
volatility matrix can always be transformed to a positive
semi-definite matrix which facilitates the optimization.

We run two investment strategies. In the first strategy, the
portfolio is held for $\tau = 1$ trading day before we re-estimate
the covariation structure and adjust the portfolio weights
accordingly. The second strategy is the same as the first one except
for the fact that the portfolio is held for $\tau = 5$ trading days
before rebalance.

In the investment horizon (which is from day $101$ to day $200$ in
this case), we record the 15-minute portfolio returns based on the
latent prices of the assets, the variation of the portfolio weights
across 50 assets, and other relevant characteristics.  While it
appears that 100 trading days is short, calculating 15-minute
returns increases the size of the relevant data for computing the
risk by a factor of 26.

We study those portfolio features for a whole range of gross
exposure constraint $c$ from $c = 1$, which stands for the
no-short-sale portfolio strategy, to $c = 3$. This is usually the
relevant range of gross exposure for investment purpose.

The standard deviations and other characteristics of the strategy
for $\tau = 1$ are presented in Table \ref{tab2} (the case $\tau =
5$ is very similar, therefore omitted). The standard deviations
represent the actual risks of the strategy. As we only optimize the
risk profile, we should not look significantly on the returns of the
optimal portfolios. They can not even be estimated accurately with
such a short investment horizon. Figures \ref{Fig4} and \ref{Fig5}
provides graphical details to these characteristics for both $\tau =
1$ and $\tau = 5$.

\ifthenelse{\boolean{flagFourFig}} {

\begin{table}[h]             
\begin{center}
\caption{\bf The out-of-sample performance of daily-rebalanced
optimal portfolios with gross-exposure constraint} \label{tab2}
\end{center}
\vspace*{-0.3 in}

\noindent \small We simulate one trial of intra-day trading data for
50 assets, make portfolio allocations for $100$ trading days and
rebalance daily. The means, standard deviations and other
characteristics of these portfolios are recorded. For all-refresh RK
approach, the bandwidth $H$ is chosen to be $1$. (All the
characteristics are annualized. Max Weight: Median of maximum
weights; Min Weight: Median of minimum weights; No. of Long: Median
of numbers of long positions whose weights exceed 0.001; No. of
Short: Median of numbers of short positions whose absolute weights
exceed 0.001)
\begin{center}
\small \begin{tabular}{lccccccc} \hline
& Mean & Std Dev  & Return-Risk & Max    & Min    & No. of &  No. of\\
Methods & \% & \% &  Ratio & Weight & Weight &   Long & Short
 \\ \hline
\multicolumn{8}{c}{\em {Low Frequency Sample Covariance Matrix Estimator}} \\
\hline
 c = 1 (No short)        & 68.90 & 16.69 & 4.13 & 0.19 & -0.00 &  13  &  0 \\
 c = 2                   & 41.91 & 16.44 & 2.55 & 0.14 & -0.05 & 28.5 & 20 \\
 c = 3                   & 41.94 & 16.45 & 2.55 & 0.14 & -0.05 & 28.5 & 20 \\
 \hline

 \multicolumn{8}{c}{\em {High Frequency All-Refresh TSRV Covariance Matrix Estimator}} \\
\hline
 c = 1 (No short)        & 54.73 & 16.08 & 3.40 & 0.20 & -0.00 & 15 &  0 \\
 c = 2                   &  9.41 & 14.44 & 0.65 & 0.14 & -0.05 & 30 & 19 \\
 c = 3                   &  9.41 & 14.44 & 0.65 & 0.14 & -0.05 & 30 & 19 \\
 \hline

 \multicolumn{8}{c}{\em {High Frequency All-Refresh RK Covariance Matrix Estimator}} \\
\hline
 c = 1 (No short)        & 42.84  & 17.20 &  2.49 & 0.22 & -0.00 & 12.5 &  0 \\
 c = 2                   & -28.94 & 20.35 & -1.42 & 0.22 & -0.09 & 22   & 18 \\
 c = 3                   & -62.06 & 31.37 & -1.98 & 0.34 & -0.19 & 23.5 & 23 \\
 \hline

\multicolumn{8}{c}{\em {High Frequency Pairwise-Refresh TSRV Covariance Matrix Estimator}}\\
\hline
 c = 1 (No short)        & 53.07 & 15.34 & 3.46 & 0.18 & -0.00 & 15 &  0 \\
 c = 2                   &  7.38 & 12.72 & 0.58 & 0.13 & -0.03 & 31 & 18 \\
 c = 3                   &  7.38 & 12.72 & 0.58 & 0.13 & -0.03 & 31 & 18 \\
 \hline

\hline
\end{tabular}
\end{center}
\end{table}

\begin{figure}[t]   
\begin{center}
\begin{tabular}{c c}
\includegraphics[scale=0.7]{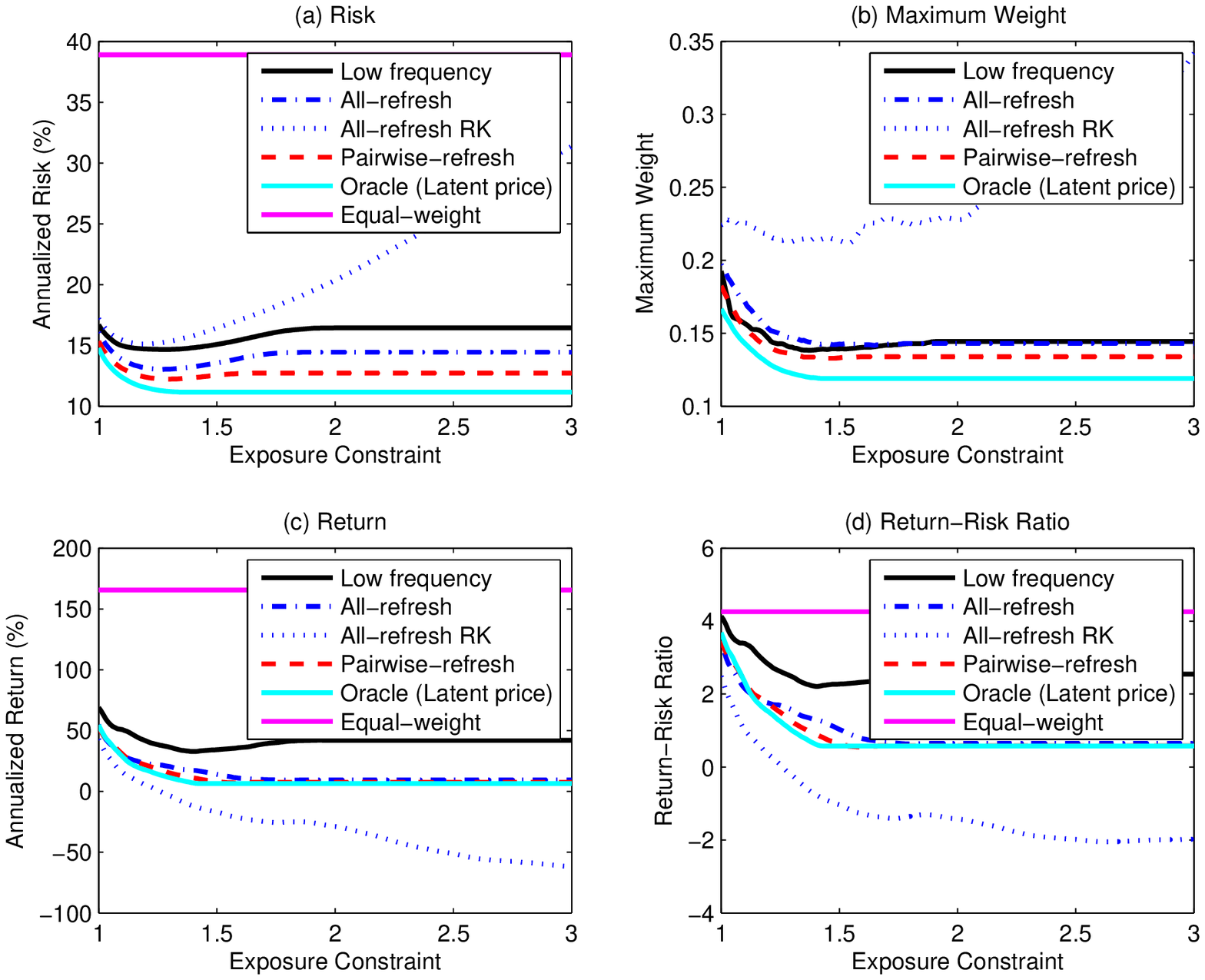} &
\end{tabular}
   \caption{Out-of-sample performance of daily-rebalanced optimal portfolios based on high-frequency and low-frequency estimation of the integrated covariance matrix.
    (a) Annualized risk of portfolios. (b) Maximum weight of allocations.
    (c) Annualized return of portfolios. (d) Return-risk ratio of portfolios
    } \label{Fig4}
\end{center}
\end{figure}

\begin{figure}[t]   
\begin{center}
\begin{tabular}{c c}
\includegraphics[scale=0.7]{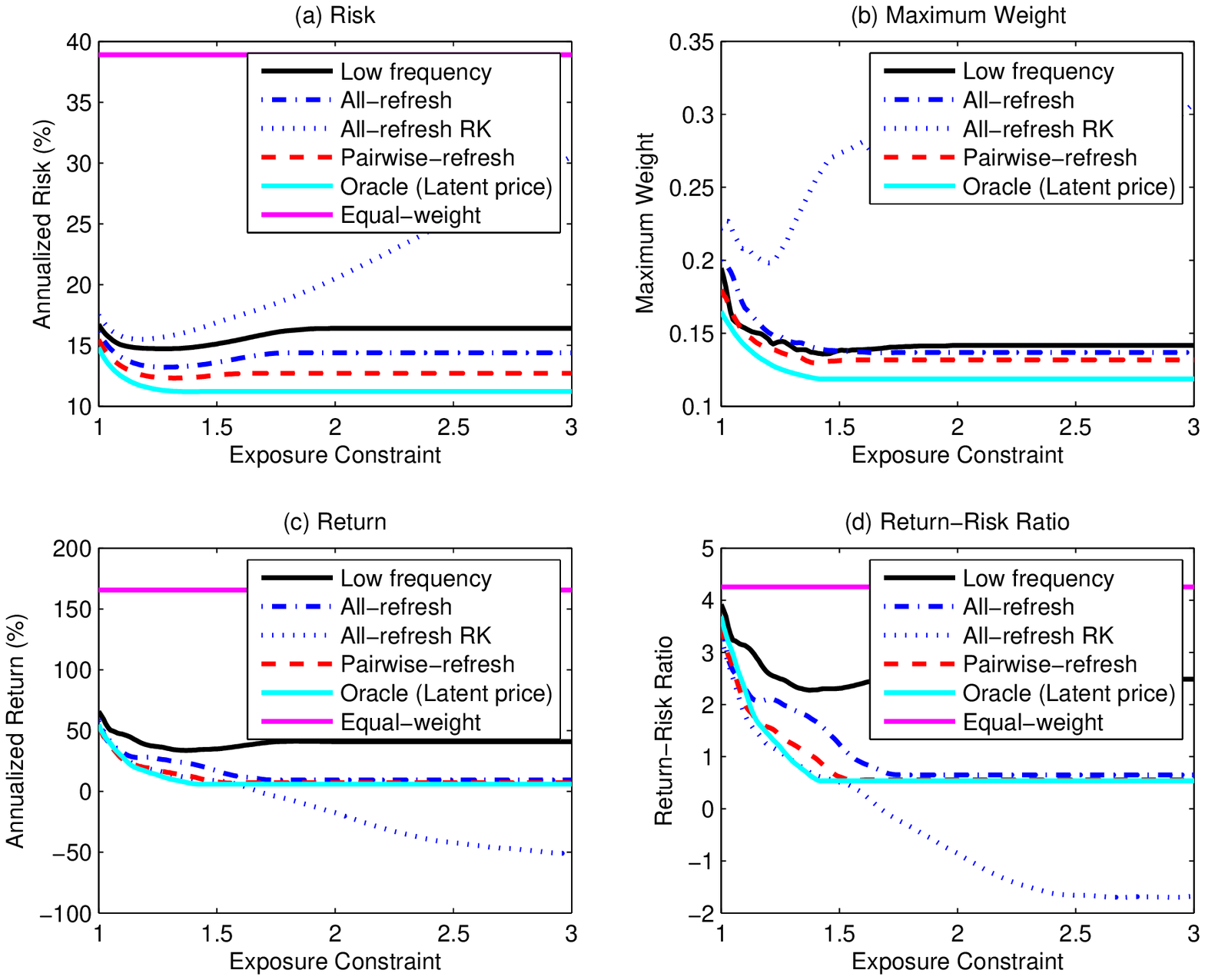} &
\end{tabular}
   \caption{Out-of-sample performance of 5-day-rebalanced optimal portfolios based on high-frequency and low-frequency estimation of the integrated covariance matrix.
    (a) Annualized risk of portfolios. (b) Maximum weight of allocations.
    (c) Annualized return of portfolios. (d) Return-risk ratio of portfolios
    } \label{Fig5}
\end{center}
\end{figure}

} {

\begin{table}[h]             
\begin{center}
\caption{\bf The out-of-sample performance of daily-rebalanced
optimal portfolios with gross-exposure constraint} \label{tab2}
\end{center}
\vspace*{-0.3 in}

\noindent \small We simulate one trial of intra-day trading data for
50 assets, make portfolio allocations for $100$ trading days and
rebalance daily. The standard deviations and other characteristics
of these portfolios are recorded. All the characteristics are
annualized (Max Weight: Median of maximum weights; Min Weight:
Median of minimum weights; No. of Long: Median of numbers of long
positions whose weights exceed 0.001; No. of Short: Median of
numbers of short positions whose absolute weights exceed 0.001)
\begin{center}
\small \begin{tabular}{lccccc} \hline
& Std Dev  & Max    & Min    & No. of &  No. of\\
Methods & \% & Weight & Weight &   Long & Short
 \\ \hline
\multicolumn{6}{c}{\em {Low Frequency Sample Covariance Matrix Estimator}} \\
\hline
 c = 1 (No short)        & 16.69 & 0.19 & -0.00 &  13  &  0 \\
 c = 2                   & 16.44 & 0.14 & -0.05 & 28.5 & 20 \\
 c = 3                   & 16.45 & 0.14 & -0.05 & 28.5 & 20 \\
 \hline

 \multicolumn{6}{c}{\em {High Frequency All-Refresh TSRV Covariance Matrix Estimator}} \\
\hline
 c = 1 (No short)        & 16.08 & 0.20 & -0.00 & 15 &  0 \\
 c = 2                   & 14.44 & 0.14 & -0.05 & 30 & 19 \\
 c = 3                   & 14.44 & 0.14 & -0.05 & 30 & 19 \\
 \hline

 \multicolumn{6}{c}{\em {High Frequency All-Refresh RK Covariance Matrix Estimator}} \\
\hline
 c = 1 (No short)        & 17.20 & 0.22 & -0.00 & 12.5 &  0 \\
 c = 2                   & 20.35 & 0.22 & -0.09 & 22   & 18 \\
 c = 3                   & 31.37 & 0.34 & -0.19 & 23.5 & 23 \\
 \hline

\multicolumn{6}{c}{\em {High Frequency Pairwise-Refresh TSRV Covariance Matrix Estimator}}\\
\hline
 c = 1 (No short)        & 15.34 & 0.18 & -0.00 & 15 &  0 \\
 c = 2                   & 12.72 & 0.13 & -0.03 & 31 & 18 \\
 c = 3                   & 12.72 & 0.13 & -0.03 & 31 & 18 \\
 \hline

\hline
\end{tabular}
\end{center}
\end{table}

\begin{figure}[t]   
\begin{center}
\begin{tabular}{c c}
\includegraphics[scale=0.7]{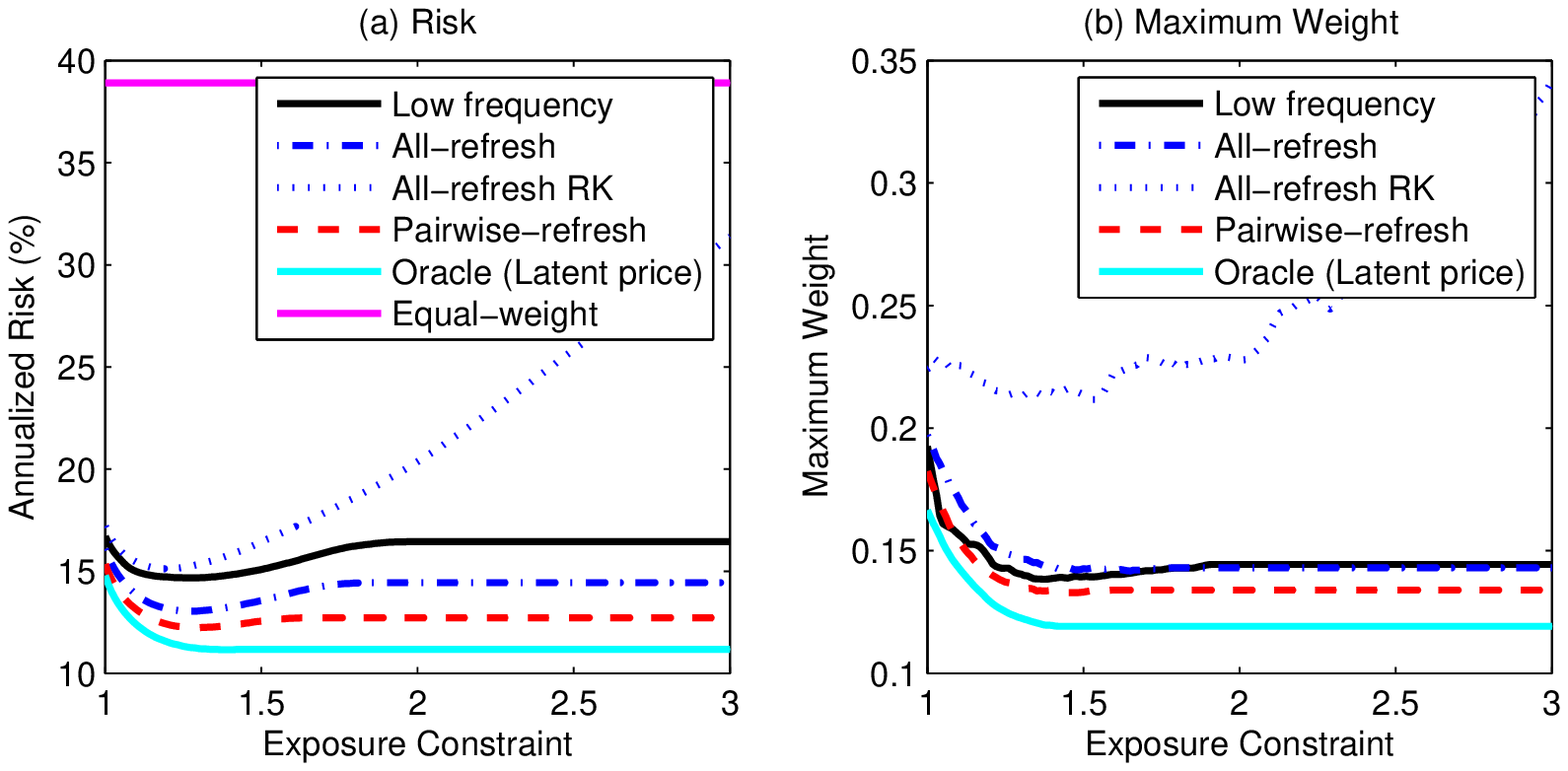} &
\end{tabular}
   \caption{Out-of-sample performance of daily-rebalanced optimal portfolios based on high-frequency and low-frequency estimation of the integrated covariance matrix.
    (a) Annualized risk of portfolios. (b) Maximum weight of allocations.
    } \label{Fig4}
\end{center}
\end{figure}

\begin{figure}[t]   
\begin{center}
\begin{tabular}{c c}
\includegraphics[scale=0.7]{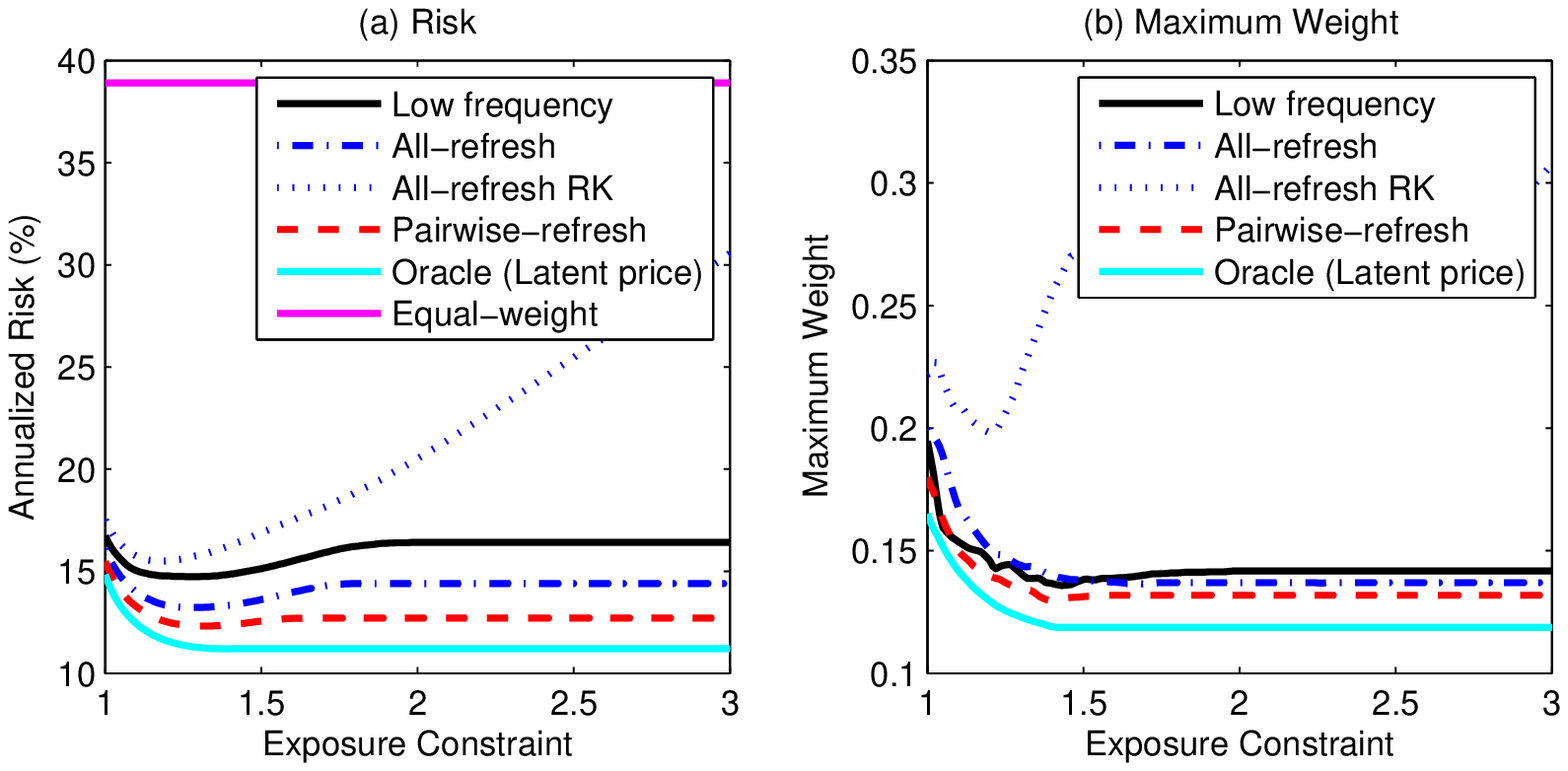} &
\end{tabular}
   \caption{Out-of-sample performance of  optimal portfolios based on high-frequency and low-frequency estimation of the integrated covariance matrix with holding period $\tau = 5$.
    } \label{Fig5}
\end{center}
\end{figure}
}


For both holding lengths $\tau = 1$ and $\tau =5$, the all-refresh
TSRV and pairwise-refresh TSRV approaches outperform significantly
the low frequency one in terms of risk profile for the whole range
of the gross exposure constraint. This supports our theoretical
results and intuitions. The shorter estimation window allows these 2
high frequency approaches to deliver consistently better results
than the low frequency one. The low-frequency strategy outperforms
significantly the equal-weight portfolio (see Figure~\ref{Fig4} and
Figure~\ref{Fig5}). Slightly surprising is the fact that the low
frequency approach also outperforms the all-refresh RK approach. We
believe it must be due to the instability of the estimated realized
kernel covariance matrix.

All the risk curves attain their minimum around $c = 1.2$ (see
Figure~\ref{Fig4} and Figure~\ref{Fig5}), which falls into our
expectation again, since that must be the point where the marginal
increase in estimation error outpaces the marginal decrease in
specification error. This, coupled with the result we get in the
empirical studies section, will give us some guidelines about what
gross exposure constraint to use in investment practice.



Firstly, the pairwise method outperforms the all-refresh method, as
expected. Secondly, the risk of the low frequency approach only
increases at a mild speed as the gross exposure constraint
increases. A possible explanation is that only $50$ assets is
considered, therefore the estimation error accumulation effect is
not dominating as badly as we were afraid it would be, given the low
frequency covariance sampling window is the previous $100$ trading
days. Another possible reason could be that as the data is generated
by a stationary stochastic model, the low frequency approach may be
able to capture some of the stationarity within the model.

In terms of portfolio weights, neither the low frequency nor the
high frequency optimal no-short-sale portfolios are well diversified
with all approaches assigning a concentrated weight of around $20\%$
to one individual asset. Their portfolio risks can be improved by
relaxing the gross-exposure constraint (Figure~\ref{Fig4} and
Figure~\ref{Fig5}).

\section{Empirical Studies}\label{sec5}

The risk minimization problem (\ref{b6}) has important applications
in asset allocation. We demonstrate its application in the stock
portfolio investment in the $30$ Dow Jones Industrial Average (DJIA)
constituent stocks (will be called the $30$ DJIA stocks for short).

The Dow Jones Industrial Average is one of the several stock market
indices created by Charles Dow, the editor of Wall Street Journal
and a co-founder of Dow Jones and Company. It is an index that shows
how 30 large, publicly-owned companies based in the United States
have traded during a standard trading session in the stock market.
We make the portfolio allocation to the constituents of the index as
of Sep 30, 2008 (The individual components of the DJIA are
occasionally changed as market conditions warrant.)



\begin{figure}[t]   
\begin{center}
\begin{tabular}{c c}
\includegraphics[scale=0.7]{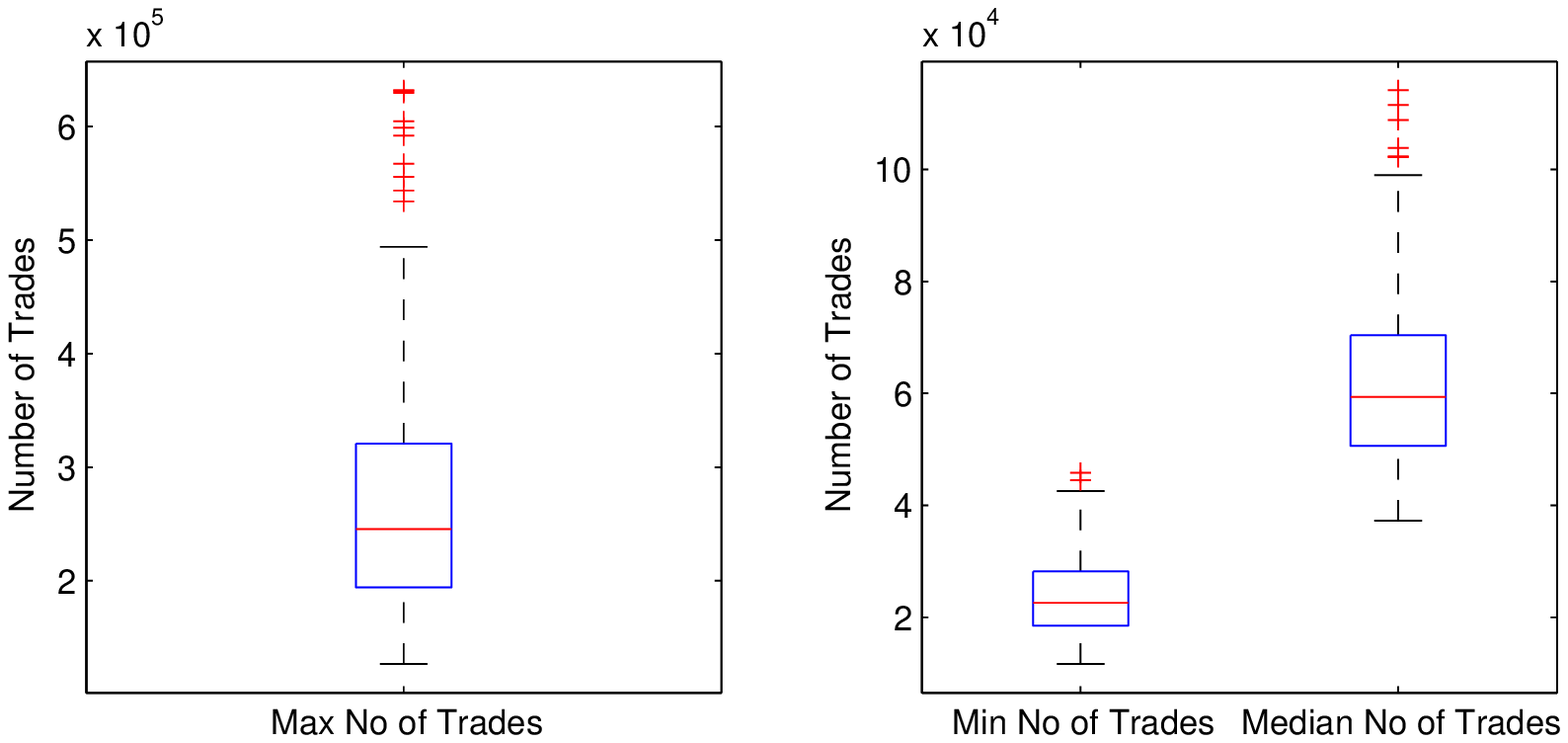} &
\end{tabular}
   \caption{The distributions (from left to right) of the maximum, minimum and median number of trades of the 30 DJIA stocks per day, from Jan 02, 2008 to Sep 30, 2008 (189 trading
   days).}\label{Fig6}
\end{center}
\end{figure}

To make asset allocation, we use the high frequency data of the $30$
DJIA stocks from Jan 1, 2008 to September 30, 2008. These stocks are
highly liquid. The intensity of trading for each given trading day
is summarized by the maximum, minimum and median number of trades
among these 30 stocks.  The distributions of these summary
statistics across those 9 months (189 trading days) are summarized
in Figure \ref{Fig6}.  The period covers the birth of financial
crisis in 2008.

At the end of each holding period of $\tau = 1$ or $\tau = 5$
trading days in the investment period (from May 27, 2008 to Sep 30,
2008), the covariance of the $30$ stocks is estimated according to
various estimators. They are the sample covariance of the last $100$
trading days' daily return data (low-frequency), the all-refresh
TSCV estimator of the last $10$ trading days, the all-refresh RK
estimator of the last $10$ trading days (the bandwidth of the
realized kernel $H$ is chosen to be 1 since the risk profile for $H
= 1$ outperforms other alternative choices of $H$), and the
pairwise-refresh TSCV estimator of the last $10$ trading days. These
estimated covariance matrices are used to construct optimal
portfolios with various exposure constraints.
For $\tau = 5$, we do not count the overnight risks of the
portifolio. The reason that the overnight price jumps are often due
to the arrival of news and are irrelevant of the topic of our
studies. The standard deviations and other characteristics of these
portfolio returns for $\tau = 1$ are presented in Table \ref{Tab3}
together with the characteristics of an equally weighted portfolio
of the $30$ DJIA stocks rebalanced daily. The standard deviations
represent the actual risks. The risk is computed based on the 15
minutes returns. Figure \ref{Fig7} and Figure \ref{Fig8} provide the
graphical details to these characteristics for both $\tau = 1$ and
$\tau = 5$.

\ifthenelse{\boolean{flagFourFig}} {

\begin{table}[h]             
\begin{center}
\caption{\bf  The out-of-sample performance of daily-rebalanced
optimal portfolios of the $30$ DJIA stocks}\label{Tab3}
\end{center}
\vspace*{-0.3 in}

\begin{center}
\small \begin{tabular}{lccccccc} \hline
& Mean & Std Dev  & Return-Risk & Max    & Min    & No. of &  No. of\\
Methods & \% & \% &  Ratio & Weight & Weight &   Long & Short
 \\ \hline
\multicolumn{8}{c}{\em {Low Frequency Sample Covariance Matrix Estimator}} \\
\hline
 c = 1 (No short)        &  0.03 & 12.73 &  0.00 & 0.50 & -0.00 &  8 &  0 \\
 c = 2                   & -8.31 & 14.27 & -0.58 & 0.44 & -0.12 & 16 & 10 \\
 c = 3                   & -0.98 & 15.12 & -0.06 & 0.45 & -0.18 & 18 & 12 \\
 \hline

 \multicolumn{8}{c}{\em {High Frequency All-Refresh TSCV Covariance Matrix Estimator}} \\
\hline
 c = 1 (No short)        & -15.94 & 12.55 & -1.27 & 0.40 & -0.00 &  8 &  0 \\
 c = 2                   & -15.69 & 12.36 & -1.27 & 0.36 & -0.10 & 17 & 12 \\
 c = 3                   & -16.77 & 12.50 & -1.34 & 0.36 & -0.10 & 17 & 12 \\
 \hline

 \multicolumn{8}{c}{\em {High Frequency All-Refresh RK Covariance Matrix Estimator}} \\
\hline
 c = 1 (No short)        &  -9.43 & 13.69 & -0.69 & 0.22 & -0.00 & 14 &  0 \\
 c = 2                   & -10.84 & 14.54 & -0.75 & 0.25 & -0.15 & 17 & 10 \\
 c = 3                   &  -6.20 & 16.55 & -0.37 & 0.30 & -0.23 & 17 & 11 \\
 \hline

\multicolumn{8}{c}{\em {High Frequency Pairwise-Refresh TSCV Covariance Matrix Estimator}}\\
\hline
 c = 1 (No short)        & -17.92 & 12.54 & -1.43 & 0.39 & -0.00 &  9 &  0 \\
 c = 2                   & -17.12 & 12.23 & -1.40 & 0.35 & -0.08 & 17 & 12 \\
 c = 3                   & -16.78 & 12.34 & -1.36 & 0.35 & -0.08 & 17 & 12 \\
 \hline

\multicolumn{8}{c}{\em {Unmanaged Index}} \\ \hline
 Dow Jones 30 equally weighted & -9.94 & 22.12 & -0.45 &&&& \\

\hline
\end{tabular}
\end{center}
\end{table}

\begin{figure}[t]   
\begin{center}
\begin{tabular}{c c}
\includegraphics[scale=0.7]{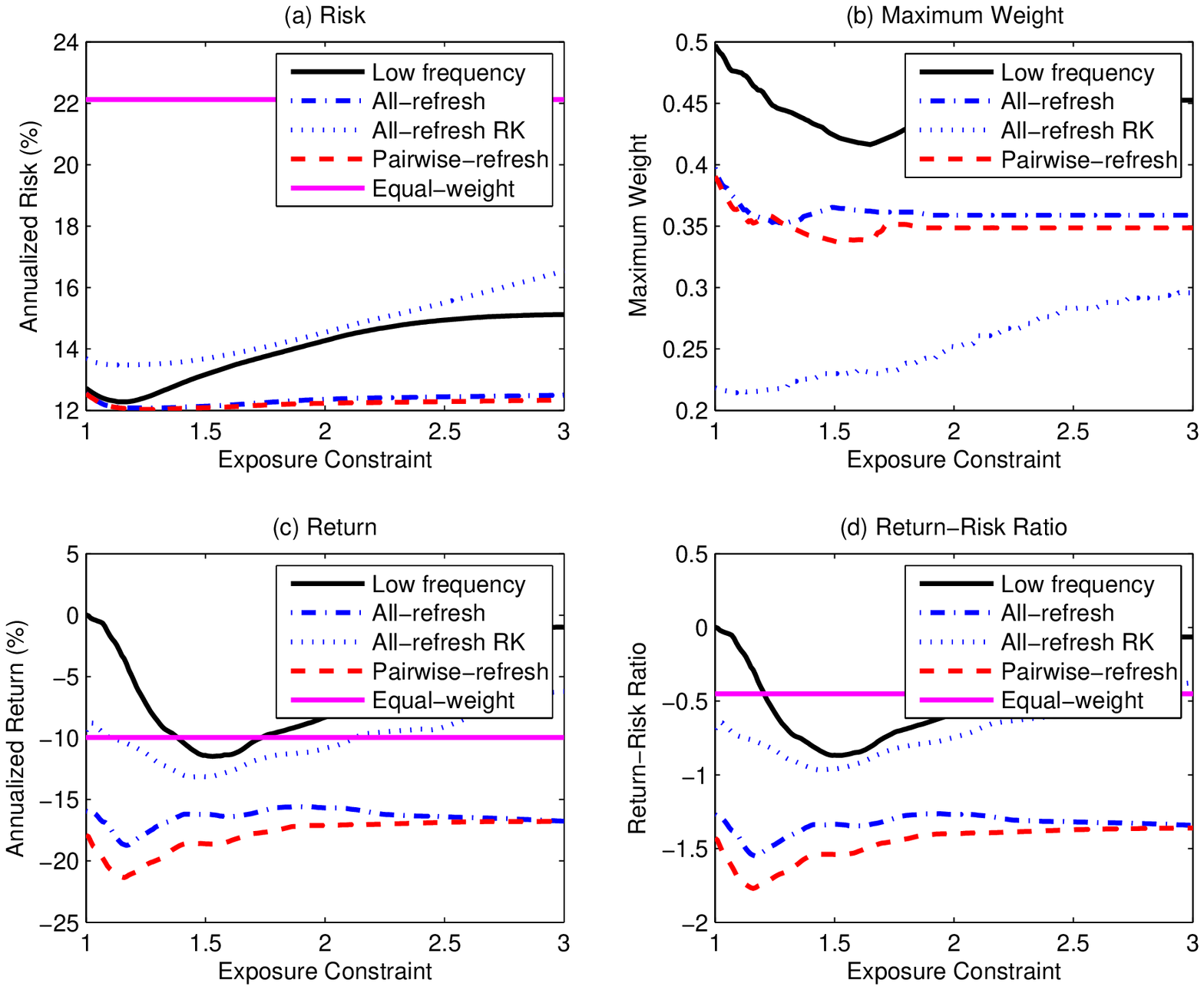} &
\end{tabular}
   \caption{Out-of-sample performance of daily-rebalanced optimal portfolios for Dow Jones $30$ constituent stocks with investment period from May 27, 2008 to Sep 30, 2008 ($89$ trading days).
    (a) Annualized risk of portfolios. (b) Maximum weight of allocations.
    (c) Annualized return of portfolios. (d) Return-risk ratio of portfolios
    } \label{Fig7}
\end{center}
\end{figure}

\begin{figure}[t]   
\begin{center}
\begin{tabular}{c c}
\includegraphics[scale=0.7]{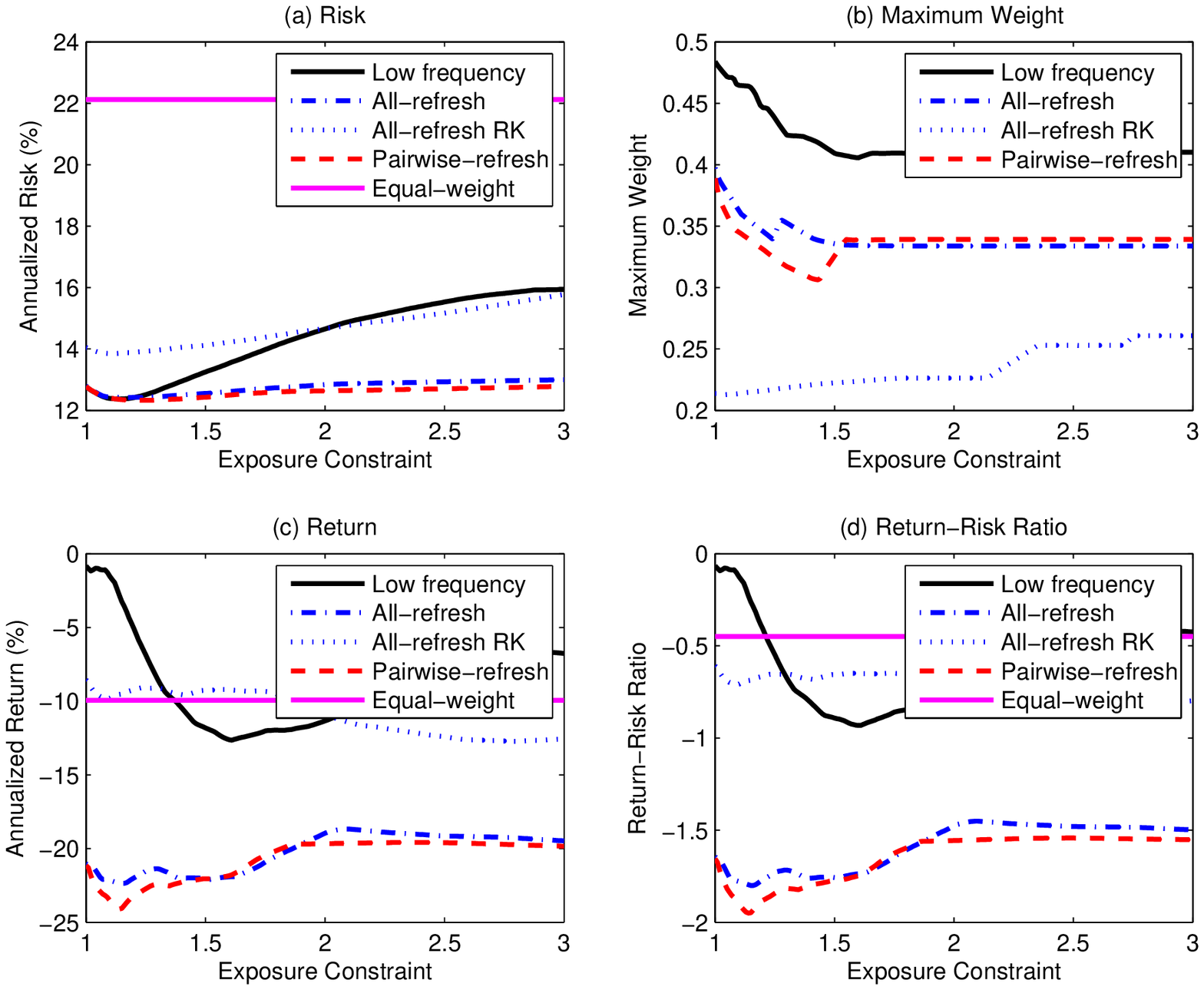} &
\end{tabular}
   \caption{Out-of-sample performance of 5-day-rebalanced optimal portfolios for Dow Jones $30$ constituent stocks with investment period from May 27, 2008 to Sep 30, 2008 ($89$ trading days).
    (a) Annualized risk of portfolios. (b) Maximum weight of allocations.
    (c) Annualized return of portfolios. (d) Return-risk ratio of portfolios
    } \label{Fig8}
\end{center}
\end{figure}

} {

\begin{table}[h]             
\begin{center}
\caption{\bf  The out-of-sample performance of daily-rebalanced
optimal portfolios of the $30$ DJIA stocks}\label{Tab3}
\end{center}
\vspace*{-0.3 in}

\begin{center}
\small \begin{tabular}{lccccc} \hline
& Std Dev  & Max    & Min    & No. of &  No. of\\
Methods & \% & Weight & Weight &   Long & Short
 \\ \hline
\multicolumn{6}{c}{\em {Low Frequency Sample Covariance Matrix Estimator}} \\
\hline
 c = 1 (No short)        & 12.73 & 0.50 & -0.00 &  8 &  0 \\
 c = 2                   & 14.27 & 0.44 & -0.12 & 16 & 10 \\
 c = 3                   & 15.12 & 0.45 & -0.18 & 18 & 12 \\
 \hline

 \multicolumn{6}{c}{\em {High Frequency All-Refresh TSCV Covariance Matrix Estimator}} \\
\hline
 c = 1 (No short)        & 12.55 & 0.40 & -0.00 &  8 &  0 \\
 c = 2                   & 12.36 & 0.36 & -0.10 & 17 & 12 \\
 c = 3                   & 12.50 & 0.36 & -0.10 & 17 & 12 \\
 \hline

 \multicolumn{6}{c}{\em {High Frequency All-Refresh RK Covariance Matrix Estimator}} \\
\hline
 c = 1 (No short)        & 13.69 & 0.22 & -0.00 & 14 &  0 \\
 c = 2                   & 14.54 & 0.25 & -0.15 & 17 & 10 \\
 c = 3                   & 16.55 & 0.30 & -0.23 & 17 & 11 \\
 \hline

\multicolumn{6}{c}{\em {High Frequency Pairwise-Refresh TSCV Covariance Matrix Estimator}}\\
\hline
 c = 1 (No short)        & 12.54 & 0.39 & -0.00 &  9 &  0 \\
 c = 2                   & 12.23 & 0.35 & -0.08 & 17 & 12 \\
 c = 3                   & 12.34 & 0.35 & -0.08 & 17 & 12 \\
 \hline

\multicolumn{6}{c}{\em {Unmanaged Index}} \\ \hline
 Dow Jones 30 equally weighted & 22.12 &&&& \\

\hline
\end{tabular}
\end{center}
\end{table}

\begin{figure}[t]   
\begin{center}
\begin{tabular}{c c}
\includegraphics[scale=0.7]{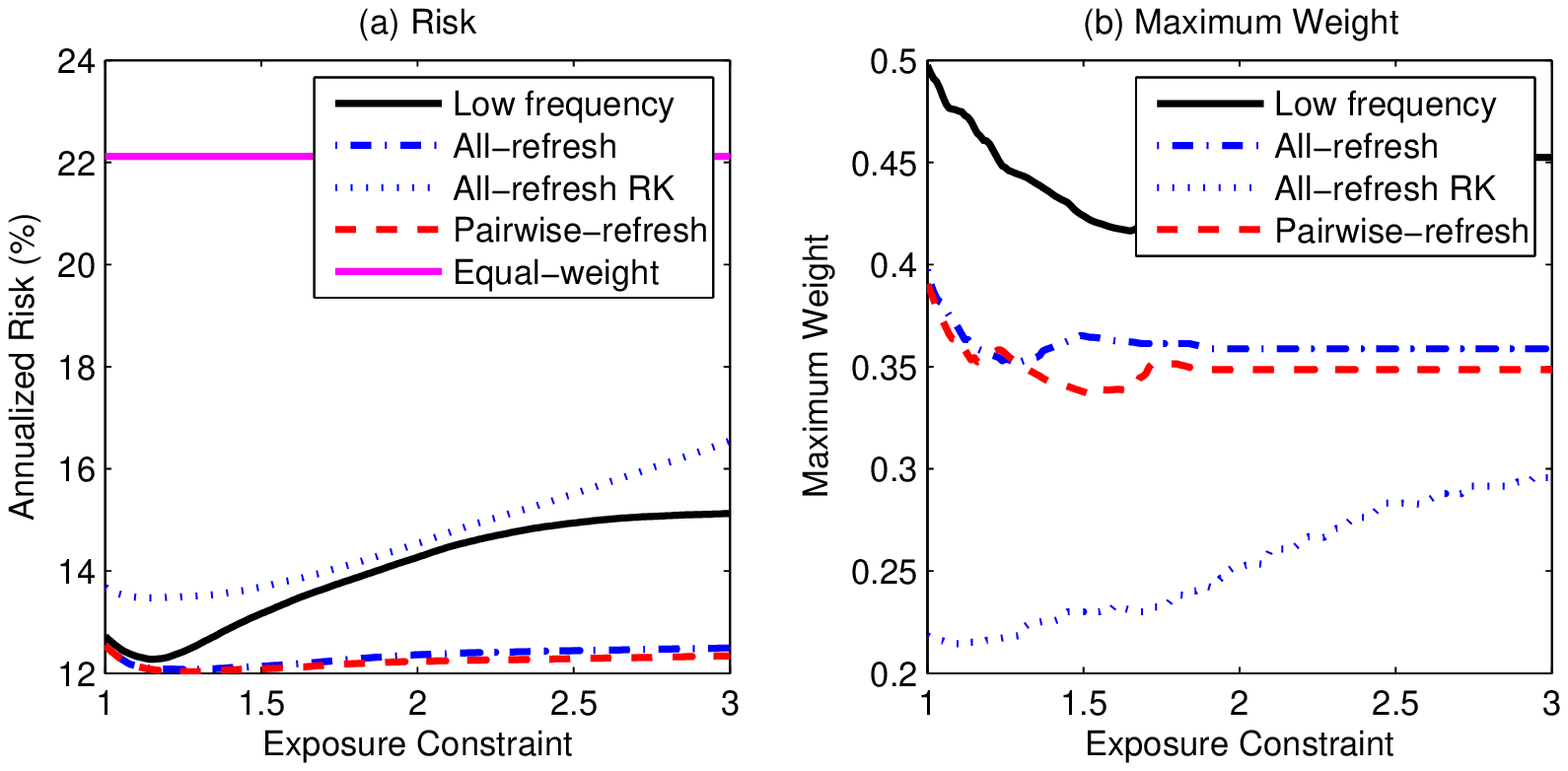} &
\end{tabular}
   \caption{Out-of-sample performance of daily-rebalanced optimal portfolios for Dow Jones $30$ constituent stocks with investment period from May 27, 2008 to Sep 30, 2008 ($89$ trading days).
    (a) Annualized risk of portfolios. (b) Maximum weight of allocations.
    } \label{Fig7}
\end{center}
\end{figure}

\begin{figure}[t]   
\begin{center}
\begin{tabular}{c c}
\includegraphics[scale=0.7]{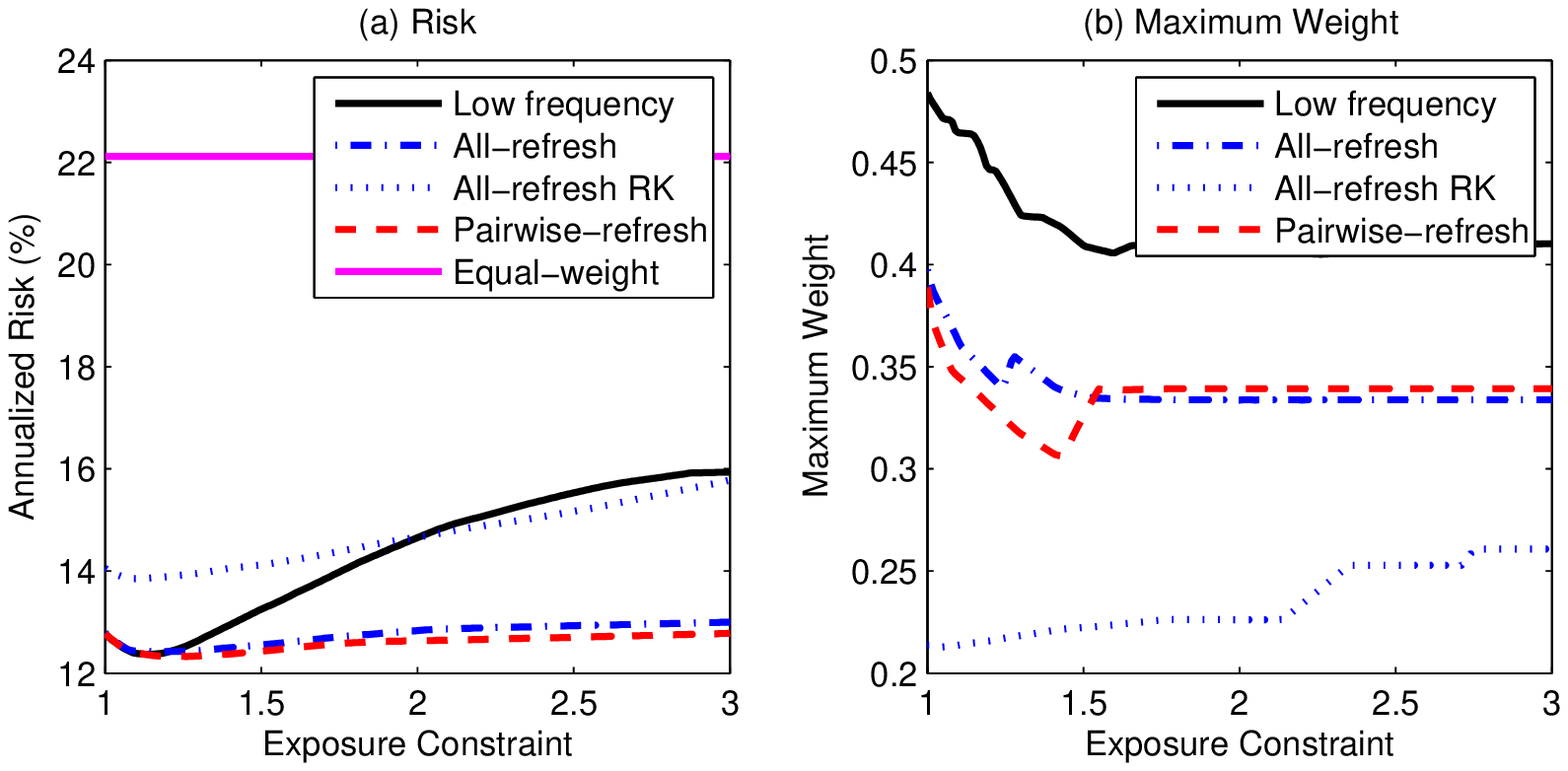} &
\end{tabular}
   \caption{Out-of-sample performance of 5-day-rebalanced optimal portfolios for Dow Jones $30$ constituent stocks with investment period from May 27, 2008 to Sep 30, 2008 ($89$ trading days).
    (a) Annualized risk of portfolios. (b) Maximum weight of allocations.
    } \label{Fig8}
\end{center}
\end{figure}

}


Table \ref{Tab3}, Figures \ref{Fig7} and \ref{Fig8} reveal that in
terms of the portfolio's actual risk, the all-refresh TSRV and
pairwise-refresh TSRV strategies perform at least as well as the low
frequency based strategy when the gross exposure is small and
outperform the latter significantly when the gross exposure is
large. Both facts support our theoretical results and intuitions.
Given $10$ times the length of covariance estimation window, the low
frequency approach still cannot perform better than the high
frequency TSRV approaches, which affirms our belief that the high
frequency TSRV approaches can significantly shorten the necessary
covariance estimation window and capture better the short-term
time-varying covariation structure (or the ``local" covariance).
These results, together with the ones presented in the simulation
section, lend strong support to the above statement.

Again the fact that the all-refresh RK strategy is outperformed by
the low frequency strategy could be due to the instability of the
estimated realized kernel covariance matrix.

As the gross exposure constraint increases, the portfolio risk of
the low frequency approach increases drastically relative to the
ones of the high frequency TSRV approaches. The reason could be a
combination of the fact that the low frequency approach does not
produce a well-conditioned estimated covariance due to the lack of
data and the fact that the low frequency approach can only attain
the long run covariation but cannot capture well the ``local"
covariance dynamics. The portfolio risk of the high frequency TSRV
approaches increased only moderately as the gross exposure
constraint increases. From financial practitioner's standpoint, that
is also one of the comparative advantages of high frequency TSRV
approaches, which means that investors do not need to be much
concerned about the choice of the gross exposure constraint while
using the high frequency TSRV approaches.

It can be seen that both the low frequency and high frequency
optimal no-short-sale portfolios are not diversified enough. Their
risk profiles can be improved by relaxing the gross-exposure
constraint to around $c=1.2$, i.e. $10\%$ short positions and
$110\%$ long positions are allowed. The no-short-sale portfolios
under all approaches have the maximum portfolio weight of $22\%$ to
$50\%$. As the gross exposure constraint relaxes, the
pairwise-refresh TSRV approach has its maximum weight reaching the
smallest value around $30\%$ to $34\%$ while the low frequency
approach goes down to only around $40\%$. That is another
comparative advantage of the high frequency approach in practice as
a portfolio with less weight concentration is always considered more
preferable by most of the investors.

Another interesting fact is that the equally weighted
daily-rebalanced portfolio of the $30$ DJIA stocks carries an
annualized return of only $-10\%$ while DJIA went down 13.5\% during
the same period (May 27, 2008 to Sep 30, 2008), giving an annualized
return of -38.3\%. The cause of the difference is that we
intentionally avoided holding portfolios overnight, hence not
affected by the overnight price jumps. In the turbulent financial
market of May to September 2008, that means our portfolio strategies
are not affected by the numerous sizeable downward jumps. Those
jumps are mainly caused by the news of distressed economy and
corporations. The moves could deviate far from what the previously
held covariation structure dictates.

\section{Conclusion} \label{sec6}

We advocate the portfolio selection with gross-exposure constraint
\citep{FanZY08}. It is less sensitive to the error of covariance
estimation and is immune to the noise accumulation. The
out-of-sample portfolio performance depends on the expected
volatility in the holding period. It is at best approximated and the
gross-exposure constraints help reducing the error accumulation in
the approximations.

Two approaches are proposed for the use of high-frequency data to
estimate the integrated covariance:  ``all-refresh" and
``pairwise-refresh" methods. The latter retains far more data on
average and hence estimates more precisely element by element. Yet,
the pairwise-refresh estimates are typically not positive
semi-definite and projections are needed for the convex optimization
algorithms. The projection distorts somewhat the performance of the
pairwise-refresh strategies.

The use of high frequency financial data increases significantly the
available sample size for volatility estimation, and hence shortens
the time window for estimation, adapts better to local covariations.
Our theoretical observations are supported by the empirical studies
and simulations, in which we demonstrate convincingly that the
high-frequency based strategies outperform the low-frequency based
one in general.

With the gross-exposure constraint, the impact of the size of the
candidate pool for portfolio allocation is limited. We derive the
concentration inequalities to demonstrate this theoretically.
Simulation and empirical studies also lend further support to it.

\appendix

\section{APPENDIX. Conditions and Proofs}

\subsection{Conditions \label{secA.1}}

The following conditions are needed.  For simplicity, we state the
conditions for integrated covariation (Theorem ~\ref{thm2}). The
conditions for integrated volatility (Theorem~\ref{thm1}) are simply
the ones with $Y = X$.

\noindent{\bf Condition 1.} $\mu_t^{(X)}=\mu_t^{(Y)}=0$.


\noindent {\bf Condition 2.} $0 < \sigma_t^{(X)}, \;
\sigma_t^{(Y)}\leq C_\sigma<\infty$, $\forall t \in [0, 1]$.


\noindent  {\bf Condition 3.} The observation times are independent
with the  $X$ and $Y$ processes.  The synchronized observation times
for the $X$ and $Y$ processes satisfy  $\sup_{1\leq j\leq \tilde
n}\tilde n\cdot(v_j-v_{j-1})\leq C_\Delta\leq \infty$,  where
${\tilde n}$ is the observation frequency and
$\mathcal{V}=\{v_0,v_1,\cdots,v_{\tilde n}\}$ is the set of refresh
times of the processes $X$ and $Y$.

 \noindent  {\bf Condition 4.}
For the TSCV parameters, we consider the case when $J=1$
($\bar{n}_J=\tilde n$) and $\bar{n}_K = O(\tilde n^{1/3})$ such that
$$ \frac{1}{2}\cdot {\tilde n}^{1/3} \leq \bar{n}_K \leq 2 \cdot {\tilde n}^{1/3}
$$

\noindent  {\bf Condition 5.} The processes $\epsilon^{X}$ and
$\epsilon^{Y}$ are independent.

Conditions 1 and 4 are imposed for simplicity.  They can be removed
at the expenses of lengthier proofs.  For a short horizon and
high-frequency, whether Condition 1 holds or not has little impact
on the investment. For estimating integrated volatility, the
synchronized time becomes observation time $\{\tau_{n, j}\}$ and
Condition 3 and 5 becomes
\begin{equation} \label{b26}
  \sup_{1\leq j\leq n} n \cdot(\tau_{n,j}-\tau_{n,j-1})\leq C_\Delta < \infty
\end{equation}
and \begin{equation} \nonumber
  \frac{1}{2}\cdot {n}^{1/3} \leq \bar{n}_K \leq 2 \cdot {n}^{1/3}.
\end{equation}

\subsection{Lemmas}

We need the following three lemmas for the proof of
Theorems~\ref{thm1} and \ref{thm2}.  In particular,
Lemma~\ref{lemma2} is exponential type of inequality for any
dependent random variables that have a finite moment generation
function. It is useful for many statistical learning problems.
Lemma~\ref{lemma3} is a concentration inequality for the realized
volatility based on discretely observed latent process.

{\lemma \label{lemma1} When $Z\sim N(0,1)$, for any $|\theta| \leq
\frac{1}{4}$,
$$E\exp\{\theta(Z^2-1)\}\leq \exp(2\theta^2).$$
}

{\bf Proof.} Using the moment generating function of $Z^2\sim
\chi^2_1$, we have
$$E\exp\{\theta(Z^2-1)\}=\exp\{-\frac{1}{2}\log(1-2\theta)-\theta\}.$$
Let $g(x)=\log(1-x)+x+x^2$ with $|x|\leq 1/2$. Then,
$g'(x)={x(1-2x)}/({1-x})$ is nonegative when $x \in [0, 1/2]$ and
negative when $x \in[-1/2,0)$.
 In other words, $g(x)$ has a minimum at point $0$, namely
$g(x) \geq 0$ for $|x|\leq 1/2$. Consequently, for $|\theta|\leq
1/4$,
$$
    \log(1-2\theta)\geq -2\theta - (2\theta)^2.
$$
Hence,
$$
        E\exp\{\theta(Z^2-1)\}\leq \exp(2\theta^2).
$$

{\lemma \label{lemma2} For a set of random variables $X_i$,
$i=1,\cdots, K$, if when $|\theta| \leq C_{1}$, \begin{equation}
\label{ineq:MGF} E \exp(\theta X_i)\leq
\exp(C_2\theta^2),\end{equation} for some two positive constants
$C_{1}$ and $C_2$, then
$$P\{|\sum_{i=1}^K w_i X_i|>x\}\leq 4 \exp \Big (- \frac{x^2}{16 C_2 w^2} \Big ), \mbox{ when }0  \leq x\leq2C_{1}C_2, $$
where $w_i$'s are weights satisfying  $\sum_{i=1}^K |w_i|\leq
w\in[1,\infty)$.}

{\bf Proof.} By the Markov inequality, for $0\leq \theta\leq C_{1}$,
we have
\begin{equation} \label{b27}
    P(|X_i|>x)\leq \exp(-\theta x) E\exp(\theta | X_i|)
    \leq 2 \exp(C_2 \theta^2-\theta x).
\end{equation}
Taking $\theta= x/(2C_2)$, we have
\begin{equation}\label{b28}
    P\{|X_i|>x\}\leq 2 \exp(-\frac{x^2}{4C_2}),
    \quad \mbox{ when } 0 \leq x \leq x_0,
\end{equation}
where $x_0=2C_{1}C_2$.

For a small constant $\xi>0$ to be specified later,  let
$$
    g_\xi(x)=\Bigl \{ \begin{array}{l l}
                       \exp (\xi x^2) \; & \mbox{ when } 0\leq x \leq x_0\\
                        \exp(a_\xi+b_\xi x) \;& \mbox{ when } x \geq x_0,
                       \end{array}
$$
where $a_\xi=-\xi x_0^2$ and $b_\xi=2\xi x_0.$ Then $g_\xi(x)$ is a
continuously differentiable increasing convex function on
$[0,\infty)$. It follows from the Markov inequality and the
convexity that, for $w^* = \sum_{i = 1}^K |w_i|$
\begin{eqnarray}
     P(|\sum_{i=1}^K w_i X_i|>x)
    & \leq & g_{\xi}(x)^{-1}E g_\xi (|\sum_{i=1}^K w_iX_i|) \nonumber \\
    & \leq & g_\xi(x)^{-1}w^{*-1}\sum_{i=1}^K|w_i| E g_\xi(w|X_i|), \label{b29}
\end{eqnarray}
which is further bounded by $4g_\xi(x)^{-1}$ if we can show that 4
is a common bound for $\{E g_\xi(w|X_i|)\}$.

Note that by (\ref{b27}) for $w b_\xi \leq \theta \leq C_{1}$,
$$
    \lim_{x\to\infty}g_\xi(x)\cdot P\{w|X_i|>x\} = 0.
$$
It follows from the integration by parts that
\begin{equation} \aligned
 E g_\xi(w|X_i|) =& 1 + \int_0^{x_0}2\xi x \exp(\xi x^2) P (w|X_i|>x) dx\\& + \int_{x_0}^\infty b_\xi \exp(a_\xi + b_\xi x) P\{w|X_i|>x\}dx.
 \label{b30}       \endaligned
\end{equation}
By (\ref{b28}), the second term in (\ref{b30}) is bounded by
$$
    \int_0^{x_0} 2\xi x \exp(\xi x^2) 2 \exp(-C_3 x^2) dx
  = \frac{2 \xi }{C_3 - \xi} \Bigl ( 1 - \exp\{(\xi-C_3)x_0^2\} \Bigl ),
$$
where $C_3=(4C_2 w^2)^{-1}$. Using (\ref{b27}), the third term in
(\ref{b30}) is bounded by
\begin{eqnarray*}
 & & 2 b_\xi \exp(a_\xi)\int_{x_0}^\infty \exp(b_\xi  x + C_2{\theta}^2- \theta w^{-1} x)dx \\
  & = & 2 b_\xi\exp(a_\xi+C_2 {\theta}^2+b_\xi  x_0-{\theta} w^{-1} x_0)/
  ({\theta} w^{-1}-b_\xi),
\end{eqnarray*}
provided that $wb_\xi< {\theta}\leq C_{1}$.

Choosing further $\theta$ to satisfy $a_\xi+C_2 {\theta}^2+b_\xi
x_0-{\theta} w^{-1} x_0 \leq 0$ and $\xi \leq C_3$, it follows from
the calculation in the previous paragraph that
$$
  E(g_\xi(w|X_i|))  \leq  1+\frac{2 \xi }{C_3-\xi} + \frac{2 b_\xi}{{\theta} w^{-1}-b_\xi}.
$$
Taking ${\theta}=\theta_0=\frac{C_{1}}{2w}$
and $\xi=\xi_0=\frac{1}{16C_2w^2}$, which satisfy the above
conditions, it follows from direct calculation that
$$
    E g_{\xi_0}(w|X_i|) \leq  1+\frac{2 \xi_0 }{C_3-\xi_0}+\frac{2b_{\xi_0}}{\theta_0 w^{-1}-b_{\xi_0}} =\frac{11}{3}<4.
$$

To summarize, from the above, we know that
$$
    E g_{\xi_0}(w|X_i|)\leq 4 \mbox{ for all } i=1,\cdots, K.
$$
Therefore, continued from (\ref{b29}), we have
$$
    P(|\sum_{i=1}^K w_i X_i|>x)\leq 4 g_{\xi_0}(x)^{-1} =
    4 \exp\{-\frac{x^2}{16C_2 w^2} \} \mbox{ when } 0\leq x \leq 2C_{1}C_2.
$$
This completes the proof of the lemma.

{\lemma \label{lemma3} ({\bf A Concentration Inequality for Realized
Volatility Based on Latent Process}) For a one dimensional process
$X_t$ following (\ref{b1}) that satisfies Conditions 1-2, when one
observes $X_t$ at times $v_i$, $i=1,\cdots,n$, and the observation
frequency satisfies Condition 3 (see (\ref{b26})), then, for $x \in
[0, c \sqrt{n} \;]$,
$$
  P \Big \{n^{1/2}| \widetilde{[X,X]_1}-\int_0^1\sigma_t^2 dt|>x
  \Big \} \leq  4 \exp \{-  C x ^{2} \},$$ where $
  \widetilde{[X,X]_1}=\sum_{i=1}^{n}(X_{v_i}-X_{v_{i-1}})^2$ is the realized volatility based on the
  discretely observed $X$ process;
   the constants $c$ and $C$ can be taken as in (\ref{dev:rv}).}

{\bf Proof.}  Let $\tilde X_t = X_t - X_0 = \int_0^t \sigma_s dW_s$.
 By time-change for  martingales, (see, for example,
Theorem 4.6 of \cite{KaratzasS00}), if $ \tau_t=\inf\{s:[\tilde
X]_s\geq t\}$ where $[\tilde X]_s$ is the quadratic variation
process, then $ B_t:=\tilde X_{\tau_t} $ is a Brownian-motion w.r.t.
$\{\mathcal{F}_{\tau_t}\}_{0\leq t\leq\infty}$. We then have that
$$
    E\exp \Bigl ( \theta(\tilde X_t^2 - \int_0^t \sigma_s^2 ds) \Bigr )
         = E\exp\big(\theta(B_{[\tilde X]_t}^2 - [\tilde X]_t)\big ).
$$
Note further that for any $t$, $[\tilde X]_t$ is a stopping time
w.r.t. $\{\mathcal{F}_{\tau_s}\}_{0\leq s\leq \infty}$, and the
process $\exp \Bigl ( \theta(B_{s}^2 - s) \Bigr )$ is a
sub-martingale for any $\theta$.  By the optional sampling theorem,
using $[\tilde X]_u\leq C_{\sigma}^2 u$ (bounded stopping time), we
have
$$
  E \exp \Bigl ( \theta(B_{[\tilde X]_u}^2 - [\tilde X]_u) \Bigr )
  \leq E \exp \Bigl (\theta(B_{ C_{\sigma}^2 u}^2 -  C_{\sigma}^2 u)
     \Bigr ).
$$
Therefore, we have that, under Conditions 2 and 3,
\begin{eqnarray}  \label{exp:dx}
& & \notag  E \Bigl (\exp \Bigl \{ \theta\sqrt{n } ((\Delta X_i)^2
  - \int_{v_{i-1}}^{v_{i}} \sigma_t^2 dt)\Bigr \}|\mathcal{F}_{v_{i-1}}\Bigl )  \\
 & \leq & E \exp \Bigl \{\theta \sqrt{n}(B^2_{\frac{C_\sigma^2C_\Delta}{n}} -
 \frac{C_\sigma^2C_\Delta}{n}) \Bigr \} \nonumber \\
& = & E\exp \Bigl \{\theta
\frac{C_{\sigma}^2C_\Delta}{\sqrt{n}}(Z^2-1)
    \Bigr \},
\end{eqnarray}
where $Z\sim N(0,1)$ and $\Delta X_i = X_{v_{i}} - X_{v_{i-1}}$.

It follows from the law of iterated expectations and (\ref{exp:dx})
that
\begin{eqnarray*}
 &   & E  \exp\bigr \{ \theta\sqrt{{n}}( \widetilde{[X,X]_1} -
       \int_0^1 \sigma_t^2 dt)\bigr \}\nonumber \\
 & = & E  (\exp \bigl \{ \theta\sqrt{{n}}
        (\sum_{i=1}^{n-1} (\Delta X_i)^2 - \int_0^{v_{{n}-1}}
         \sigma_t^2 dt)\bigr \} \\ & & \times E\Bigl (\exp \bigl \{\theta\sqrt{{n}}(\Delta
X_{n }^2- \int_{v_{n-1}}^{v_{n}} \sigma_t^2 dt )\bigr \}
 |\mathcal{F}_{v_{n-1}} \Bigl ))  \nonumber \\
&\leq & E  \exp \bigl \{ \theta\sqrt{n}
        (\sum_{i=1}^{n-1} (\Delta X_i)^2 - \int_0^{v_{n-1}}
         \sigma_t^2 dt)\bigr \}
E \exp \Big \{\theta \frac{ C_{\sigma}^2C_{\Delta}}{\sqrt{n}}(Z^2-1)
\Big \} ,
\end{eqnarray*}
where $Z\sim N(0,1)$.  Repeating the process above, we obtain
$$
 E  \exp\bigr \{ \theta\sqrt{{n}}( \widetilde{[X,X]_1}-
       \int_0^1 \sigma_t^2 dt)\bigr \}
  \leq \left ( E \exp \Big \{\theta \frac{C_{\sigma}^2C_{\Delta}}{\sqrt{n}}(Z^2-1) \Big \}
    \right )^{n}.
$$
By Lemma 1, we have for $|\theta|\leq \frac{\sqrt{n}}{4 C_\sigma^2
C_\Delta}$,
\begin{equation} \label{exp:rv}
 E  \exp\bigr \{ \theta\sqrt{n}( \widetilde{[X,X]_1} -
       \int_0^1 \sigma_t^2 dt)\bigr \}
   \leq \exp\{2\theta^2 C_{\sigma}^4 C_{\Delta}^2\}.
\end{equation}

By Lemma 2, we have,

\begin{equation} \label{dev:rv} P \Big \{n^{1/2}| \widetilde{[X,X]_1}-\int_0^1\sigma_t^2 dt|>x
  \Big \} \leq  4 \exp \{-  \frac{x ^{2}}{32C_\sigma^4 C_\Delta^2
  }\},
  \end{equation}
  when $ 0 \leq x \leq   C_\sigma^2
C_\Delta\sqrt{n}  $.

\subsection{Proof of Theorem 1}
We first prove the results conditional on the set of observation
times $\mathcal{V}$. Recall notation introduced in sections
\ref{sec3.2} and \ref{sec3.3}. Let $n$ be the observation frequency.
For simplicity of notation, without ambiguity, we will write
$\tau_{n, i}$ as $\tau_i$ and $\sigma_t^{(X)}$ as $\sigma_t$. Denote
the TSRV based on the unobserved latent process by
\begin{equation}
    \widetilde{\langle X,X \rangle}_1^{(K)} =   \widetilde{[X,X]}_1^{(K)}-\frac{\bar{n}_K}{\bar{n}_J}
    \widetilde{[X,X]_1}^{(J)},
\end{equation}
where $\widetilde{[X,X]}_1^{(K)} = K^{-1} \sum_{i=K}^{n}
(X_{\tau_{i}}-X_{\tau_{i-K}})^2$. Then, from the definition, we
have,
\begin{eqnarray}\label{b32}
    \widehat{\langle   X, X \rangle}_1 & = &\widetilde{[X,
    X]}_1^{(K)} + \widetilde{[\epsilon^{X}, \epsilon^{X}]}_1^{(K)}
    + 2 \widetilde{[X,\epsilon^{X}]}_1^{(K)}\notag\\ && -\frac{\bar{n}_K}{\bar{n}_J}
    \left(\widetilde{[X, X]}_1^{(J)}+\widetilde{[\epsilon^{X}, \epsilon^{X}]}_1^{(J)}  +   2\widetilde{[X,\epsilon^{X}]}_1^{(J)}\right)
    \nonumber \\
    &   =  & \frac{1}{K}\sum_{l=0}^{K-1} V_K^{(l)} - \frac{\bar{n}_K}{\bar{n}_J}\widetilde{[X, X]}_1^{(J)} +  R_1 + R_2,
\end{eqnarray}
where $R_1 =  \widetilde{[\epsilon^{X}, \epsilon^{X}]}_1^{(K)} -
\frac{\bar{n}_K}{\bar{n}_J} \widetilde{[\epsilon^{X},
\epsilon^{X}]}_1^{(1)}$, $R_2 = 2
\widetilde{[X,\epsilon^{X}]}_1^{(K)} - 2\frac{\bar{n}_K}{\bar{n}_J}
\widetilde{[X,\epsilon^{X}]}_1^{(1)}$, and
$$
    V_K^{(l)} = \sum_{j=1}^{\bar{n}_K} (X_{\tau_{jK+l}}-X_{\tau_{(j-1)K+l}})^2, \mbox{ for }
l=0,1,\cdots, K-1.
$$ Note that we have assumed that $\bar{n}_K=\frac{n-K+1}{K}$ is an
integer above, to simplify the presentation.

Recall that we consider the case when $J=1$, or $\bar{n}_J=n$. We
are interested in
\begin{eqnarray}\label{b33}
&&\sqrt{\bar{n}_K} (\widehat{\langle X, X \rangle}_1-\int_0^1
{\sigma_t}^2 dt)\notag \\
 & = & \frac{1}{K}\sum_{l=0}^{K-1} \sqrt{\bar{n}_K}(V_K^{(l)}-\int_0^1
    {\sigma_t}^2 dt)      \\
  &  & - \left(\frac{\bar{n}_K}{n}\right)^{3/2}\cdot \sqrt{n} \Bigl (\widetilde{[X,X]}_1^{(1)}
    - \int_0^1 {\sigma_t}^2 dt \Bigr )\notag \\ && - \frac{\bar{n}_K^{3/2}}{n} \int_0^1 {\sigma_t}^2 dt
+ \sqrt{\bar{n}_K}R_1 + \sqrt{\bar{n}_K}R_2. \nonumber
\end{eqnarray}
The key idea is to compute the moment generating functions for each
terms in (\ref{b33}) and then to use Lemma~\ref{lemma2} to conclude.

For the first term in (\ref{b33}), since $V_k^{(l)}$  is a realized
volatility based on discretely observed $X$ process, with
observation frequency satisfying $\sup_{1\leq i \leq \bar{n}_K}
\bar{n}_K \cdot(\tau_{iK+l}-\tau_{(i-1)K+l})\leq 
 C_\Delta$,
 we have, by (\ref{exp:rv}) in Lemma 3, for $|\theta|\leq \frac{\sqrt{\bar{n}_K}}{4
C_\sigma^2 C_\Delta}$,
\begin{equation} \label{b35}
 E  \exp\bigr \{ \theta\sqrt{{\bar{n}_K}}(V_K^{(l)} -
       \int_0^1 \sigma_t^2 dt)\bigr \}
   \leq \exp\{2\theta^2 C_{\sigma}^4 C_{\Delta}^2\}.
\end{equation}

For the second term in (\ref{b33}), we have obtained in
(\ref{exp:rv}) that
\begin{equation}\label{b36}
    E\exp\big \{\theta\sqrt{n}(\widetilde{[X, X]}_1^{(1)}-\int_0^1
    {\sigma_t}^2 dt)\}\leq \exp\{2\theta^2 C_{\sigma}^4
    C_{\Delta}^2\}, \;\mbox{ when }|\theta|\leq\frac{\sqrt{n}}{4 C_\sigma^2
    C_\Delta}.
\end{equation}

The third term in (\ref{b33}) can be ignored because it has an upper
bound that goes to zero sufficiently fast as $n$ grows:
\begin{equation} \label{b37}
 \frac{\bar{n}_K^{3/2}}{n} \int_0^1 {\sigma_t}^2 dt \leq 2^{3/2}
 C_\sigma^2/\sqrt{n},
\end{equation}
by Condition 5.

We introduce an auxiliary sequence $a_n$ that grows with $n$ in a
moderate rate to facilitate our presentation in the following. In
particular, we can set $a_n=n^{1/12}$.

 Let us now deal with $R_1$, the fourth term in
(\ref{b33}).  Note that from the definition
\begin{eqnarray} \label{b38}
& & \sqrt{\bar n_K} R_1  \nonumber \\& = &
\frac{\sqrt{\bar{n}_K}}{K} \Big \{
 \sum_{i=K}^n  (\epsilon_i - \epsilon_{i-K})^2 - \frac{n - K + 1}{n} \sum_{i=1}^n (\epsilon_i - \epsilon_{i-1})^2 \Big \} \nonumber \\
 & = & \frac{\sqrt{\bar{n}_{K}}\sqrt{n}}{K} \cdot
 \frac{2}{\sqrt{n}}\sum_{i=1}^n \epsilon_i \epsilon_{i-1}\\ && \nonumber
 -\frac{\sqrt{\bar{n}_K}\sqrt{n-K+1}}{K}\cdot\frac{2}{\sqrt{n-K+1}}
 \sum_{i=K}^n \epsilon_i \epsilon_{i-K}  \\ & & -
 \frac{\sqrt{\bar{n}_K}\sqrt{K-1}a_n}{K}\cdot\frac{1}{a_n\sqrt{K-1}}
\sum_{i=1}^{K-1}(\epsilon_i^2-\eta_X^2)\nonumber\\ & & -
 \frac{\sqrt{\bar{n}_K}\sqrt{K-1}a_n}{K}\cdot\frac{1}{a_n\sqrt{K-1}}\sum_{n-K+1}^{n-1}(\epsilon_i^2-\eta_X^2)
\nonumber\\
 & & + \frac{\sqrt{\bar{n}_K} (K-1) a_n}{K\sqrt{n}}\cdot
 \frac{1}{a_n\sqrt{n}} \sum_{i=1}^n (\epsilon_i^2 -\eta_X^2)\nonumber\\ & & +\frac{\sqrt{\bar{n}_K} (K-1)a_n}{K\sqrt{n}}\cdot
 \frac{1}{a_n\sqrt{n}} \sum_{i=0}^{n-1} (\epsilon_i^2 -\eta_X^2).\nonumber
\end{eqnarray}

  The
first two terms in (\ref{b38}) are not the sum of independent
variables. But they can be decomposed into the sum of independent
random variables and the moment generating functions can be
computed.  To simplify the argument without losing the essential
ingredient, let us focus on the first term of (\ref{b38}).  It can
be decomposed as
$$
   \sum_{i=1}^n \varepsilon_{i} \varepsilon_{i-1} =
   \sum_{\mbox{odd $i$}}  \varepsilon_{i} \varepsilon_{i-1} + \sum_{\mbox{even $i$}}  \varepsilon_{i} \varepsilon_{i-1}
$$
and the summands in each terms of the right-hand side are now
independent. Therefore, we need only to calculate the moment
generating function of $\varepsilon_{i} \varepsilon_{i-1}$.

For two independent normally distributed random variables $X\sim
N(0,\sigma_X^2)$ and $Y\sim N(0,\sigma_Y^2)$, it can easily be
computed that
$$ \aligned  E(\exp\{\theta n^{-1/2}XY\})=&
\left(\frac{1}{1-\sigma_X^2\sigma_Y^2\theta^2/n}\right)^{1/2}\\\leq
&\exp\{\sigma_X^2\sigma_Y^2\theta^2/n \} \mbox{ when } |\theta|\leq
\frac{\sqrt{n}}{\sqrt{2}\sigma_X\sigma_Y} ,
\endaligned $$
where we have used the fact that $\log(1-x)\geq -2x$ when $0\leq
x\leq \frac{1}{2}$.

 Hence, by the independence, it follows that (we assume $n$ is even to simplify the presentation)
\begin{equation} \label{b39} \aligned
E\exp \Big \{ 2\theta n^{-1/2} \sum_{\mbox{odd $i$}} \varepsilon_{i}
\varepsilon_{i-1} \Big \} &=
\left(\frac{1}{1-4\eta_X^4\theta^2/n}\right)^{n/4}\\
& \leq
\exp\{2\eta_X^4\theta^2 \}, \mbox{ when } |\theta|\leq
\frac{\sqrt{n}}{2\sqrt{2}\eta_X^2}.\endaligned
\end{equation}

 The second  term in $R_1$ works similarly and
have the same bound. For example, when $\bar{n}_K$ is even, one can
have the following decomposation
$$
\sum_{i=K}^n\epsilon_i\epsilon_{i-K} =
\sum_{j=1}^{\bar{n}_K/2}\sum_{i=2jK-K}^{2jK-1}\epsilon_i\epsilon_{i-K}
+\sum_{j=1}^{\bar{n}_K/2}\sum_{i=2jK}^{2jK+K-1}\epsilon_i\epsilon_{i-K}.
$$

The last four terms are sums of independent $\chi^2$-distributed
random variables and their moment generating functions can easily be
bounded by using Lemma 1. Taking the term $ \frac{1}{a_n\sqrt{K-1}}
\sum_{i=1}^{K-1}  (\epsilon_i^2-\eta_X^2)$ for example, we have
$$E \Big (\exp\big \{\frac{\theta}{a_n\sqrt{K-1}}
\sum_{i=1}^{K-1}(\epsilon_i^2-\eta_X^2) \big \} \Big )\leq \exp\{
2\eta_X^4 \theta^2/a_n^2\}\;\mbox{ when }|\theta|\leq
\frac{a_n\sqrt{K-1}}{4\eta_X^2}.$$

For the term $R_2$, we have,
\begin{equation}\label{b40}\aligned
\sqrt{\bar{n}_K}R_2=& \frac{2 a_n \bar{n}_K}{n} \frac{1}{a_n}\Big(
\sum_{i=1}^{n} \Delta X_{i}\epsilon_{i-1}-\sum_{i=1}^{n}\Delta
X_{i}\epsilon_{i}\Big)\\&+\frac{2}{a_n}\cdot\frac{ a_n
\sqrt{\bar{n}_K}}{K} \Big( \sum_{i=K}^{n}\Delta^{(K)}
X_{i}\epsilon_{i}-\sum_{i=K}^{n} \Delta^{(K)} X_{i}\epsilon_{i-K}
\Big ),\endaligned
\end{equation} where $\Delta
X_{i}=X_{\tau_{i}}-X_{\tau_{i-1}}$, and $\Delta^{(K)}
X_{i}=X_{\tau_{i}}-X_{\tau_{i-K}}.$ The first term above satisfies
\begin{equation}\label{b41}\aligned
 E \Big (\exp\{\frac{\theta}{a_n}
\sum_{i=1}^{n}\Delta X_{i}\epsilon_{i}\} \Big ) &= E \Big (\exp\{
\sum_{i=1}^{n}(\frac{\theta}{a_n} \Delta X_{i})^2\eta_X^2/2\}
\Big )\\
&\leq \Big ( E\big(\exp\{\theta^2\eta_X^2C_\sigma^2C_\Delta Z^2
/2n{a_n^2}
\} \big) \Big)^n\\
 & = \left(\frac{1}{1-\eta_X^2 C_{\sigma}^2
C_{\Delta}\theta^2/n{a_n^2}}\right)^{n/2}\\&\leq \exp\{\eta_X^2
C_{\sigma}^2 C_{\Delta}\theta^2/a_n^2\}, \mbox{ when } |\theta|\leq
\frac{\sqrt{n}a_n}{\sqrt{2C_{\Delta}}C_{\sigma} \eta_X},
\endaligned
\end{equation}
where in the second line we have again used the optional sampling
theorem and law of iterated expectations as in the derivations of
Lemma 3; $Z$ denotes a standard normal random variable.  The second
term in $R_2$ works similarly and has the same bound. For the third
term, we have \begin{equation} \label{ieq:R22}\aligned  &E\Big[
\exp\Big\{ \frac{ a_n \theta \sqrt{\bar{n}_K}}{K}
\sum_{i=K}^{n}\Delta^{(K)} X_{i}\epsilon_{i} \Big\}\Big] \\ =&E\Big[
E\big( \exp\Big\{ \frac{ a_n \theta \sqrt{\bar{n}_K}}{K}
\sum_{i=K}^{n}\Delta^{(K)} X_{i}\epsilon_{i} \Big\}|\mbox{X
process}\big)\Big]\\
=& E\Big[ \exp\Big\{ \frac{ a_n^2 \theta^2 \bar{n}_K}{2K^2}
\sum_{l=0}^{K-1}\sum_{j=1}^{\bar{n}_K}(\Delta^{(K)} X_{i})^2
\eta_X^2 \Big\} \Big]\\
\leq & \Pi_{l=0}^{K-1} \Big\{ E\Big[ \exp\Big\{ \frac{ a_n^2
\theta^2 \bar{n}_K \eta_X^2}{2K} \sum_{j=1}^{\bar{n}_K}(\Delta^{(K)}
X_{i})^2
 \Big\} \Big] \Big\}^{\frac{1}{K}}\\
 \leq & \Pi_{l=0}^{K-1} \Big\{ \big({1-\frac{ a_n^2
\theta^2 \eta_X^2}{K}C_\sigma^2C_\Delta}\big)^{-\bar{n}_K/2}
\Big\}^{\frac{1}{K}}\\
\leq & \exp \big\{ \frac{ a_n^2 \theta^2 \bar{n}_K
\eta_X^2}{K}C_\sigma^2C_\Delta \big\} \mbox{ when } |\theta|\leq
\frac{\sqrt{K}}{\sqrt{2C_\Delta}a_n\eta C_\sigma},
\endaligned \end{equation}
where we have used the H\"{o}lder's inequality above. The forth term
works similarly and has the same bound.

Combining the results for all the terms (\ref{b35}) --
(\ref{ieq:R22}) together, applying Lemma~\ref{lemma2} to
(\ref{b33}), we have, for the following set of parameters, the
conditions for Lemma 2 are satisfied with
$C_{1}=C_{1,x}\sqrt{\bar{n}_K}$.
\begin{equation} \label{def:Ctheta}
\aligned C_{1,x}&= \min \Big \{\frac{1}{4 C_\sigma^2 C_\Delta},
\frac{\sqrt{n/\bar{n}_K}}{2\sqrt{2}\eta_X^2},
\frac{a_n\sqrt{(K-1)/\bar{n}_K}}{4\eta_X^2},
\frac{a_n\sqrt{n/\bar{n}_K}}{\sqrt{2 C_\Delta}\eta_X C_\sigma}, \frac{\sqrt{K/\sqrt{\bar{n}_K}}}{\sqrt{2C_\Delta}a_n\eta C_\sigma} \Big \}\\
&=\frac{1}{4 C_\sigma^2 C_\Delta} \mbox{ for big enough }n
,\endaligned
\end{equation}
\begin{equation}\label{def:Cm}
\aligned C_2& =\max\{2C_\sigma^4C_\Delta^2,  2\eta_X^4,
2\eta_X^4/a_n^2, \eta_X^2 C_{\sigma}^2C_{\Delta}/a_n^2, \frac{ a_n^2
 \bar{n}_K
\eta_X^2}{K}C_\sigma^2C_\Delta\}\\
&=\max\{2C_\sigma^4C_\Delta^2,  2\eta_X^4\}  \mbox{ for big enough
}n
\\ &= 2C_\sigma^4C_\Delta^2 \mbox{ considering the values }
C_{\Delta}\geq 1, \; C_\sigma \geq \eta_X \mbox{ typically},
\endaligned\end{equation} and
$$ \aligned w  = & 14 = \lceil 2+8\sqrt{2} \rceil  \\ >&
  \underbrace{\frac{1}{K}\sum_{l=0}^{K-1}1+(\frac{\bar{n}_K}{n})^{3/2}}_{\small{\mbox{
coefficients in the first two terms of }}(\ref{b33})} \\ &+
\underbrace{\frac{4\sqrt{\bar{n}_{K}}\sqrt{n}}{K}+
\frac{2\sqrt{\bar{n}_K}a_n}{\sqrt{K}}+ \frac{2\sqrt{\bar{n}_K}a_n
}{\sqrt{n}}}_{\small{\mbox{controls coefficients in }}
 (\ref{b38})}+ \underbrace{\frac{4 a_n \bar{n}_K}{n}+\frac{4}{a_n}}_{\small{\mbox{
coefficients in }} (\ref{b40})},\endaligned $$ where the $>$ is
valid when $n$ is big enough and Condition 5 is applied.

By Lemma 2, when
 $ 0 \leq x \leq 2C_{1,x}C_2 \sqrt{\bar{n}_K}$,
\begin{equation}\nonumber
    P\{\sqrt{\bar{n}_K}|\widehat{\langle X, X \rangle}_1-\int_0^1
    {\sigma_t}^2 dt|>x\}\leq  4 \exp(-(16C_2w^2)^{-1} x^2).
 \end{equation}

By the Condition 5 again, we have

\begin{equation}\label{b48} \aligned
    P\{n^{1/6}|\widehat{\langle X, X \rangle}_1-\int_0^1
    {\sigma_t}^2 dt|>x\} \leq & P(\sqrt{\bar{n}_K} |\widehat{\langle X, X \rangle}_1-\int_0^1
    {\sigma_t}^2 dt| >  x/\sqrt{2}) \\
    \leq &  4 \exp(- C x^2),\;\;\; \mbox{ when
    }0 \leq  x \leq c n^{1/6}, \endaligned
 \end{equation}
where  \begin{equation} \label{b49} c=2C_{1,x}C_2 \mbox{ and }
C=(32C_2w^2)^{-1}.
\end{equation}
This completes the proof of the result conditional on the
observation times. Theorem 1 is proved by noting  that this
conditional result depend only on the observation frequency $n$ and
not on the locations of the observation times as long as the
Condition 3 is satisfied.

Note also that in the above proof, we have demonstrated by using a
sequence $a_n$ that goes to $\infty$ at a moderate rate that, one
can eliminate the impact of the small order terms on the choices of
the constants, as long as the terms have their moment generating
functions satisfy inequalities of form (\ref{ineq:MGF}). We will use
this technique again in the next subsection.

\subsection{Proof of Theorem 2}
We again conduct all the analysis assuming the observation times are
given. Our final result holds  because the conditional result
doesn't depend on the locations of the observation times as long as
the Condition 3 is satisfied.

 Recall notation for the observation times as introduced in
section 3.2. Define $$Z^+=X+Y\;\;\mbox{ and }\;\; Z^-=X-Y.$$ $Z^+$
and $Z^-$ are diffusion processes with bounded volatility. To see
this, let $W^+$ and $W^-$ be processes such that
$$dW^+_t=\frac{\sigma_t^{(X)}dB_t^{(X)}+\sigma_t^{(Y)}dB_t^{(Y)}}{\sqrt{(\sigma_t^{(X)})^2+(\sigma_t^{(Y)})^2+2\rho_t
\sigma_t^{(X)}\sigma_t^{(Y)}}}$$  and $$
dW^-_t=\frac{\sigma_t^{(X)}dB_t^{(X)}-\sigma_t^{(Y)}dB_t^{(Y)}}{\sqrt{(\sigma_t^{(X)})^2+(\sigma_t^{(Y)})^2-2\rho_t
\sigma_t^{(X)}\sigma_t^{(Y)}}}.$$ $W^+$ and $W^-$ are standard
Brownian motions by Levy's characterization of Brownian motion (see,
for example, Theorem 3.16 of \cite{KaratzasS00}). Write
$$\sigma_t^{Z^+}=\sqrt{(\sigma_t^{(X)})^2+(\sigma_t^{(Y)})^2+2\rho_t
\sigma_t^{(X)}\sigma_t^{(Y)}} $$ and $$\sigma_t^{Z^-}=\sqrt{(\sigma_t^{(X)})^2+(\sigma_t^{(Y)})^2-2\rho_t
\sigma_t^{(X)}\sigma_t^{(Y)}},$$ we have
$$ dZ^{+}=\sigma_t^{Z^+} dW^+_t \mbox{ and }  dZ^{-}=\sigma_t^{Z^-} dW^-_t$$
with $0 \leq \sigma_t^{Z^+}, \sigma_t^{Z^-} \leq 2C_\sigma$.

For the observed $Z^+$ and $Z^-$ processes, we have
$$Z_{v_i}^{+,o}
=X_{t_i}^o+Y_{s_i}^o=Z_{v_i}^{+}+\epsilon_{i,+} \mbox{ and }
 Z_{v_i}^{-,o}=X_{t_i}^o-Y_{s_i}^o=Z_{v_i}^{-}+\epsilon_{i,-}
,$$
 where $t_i$ and $s_i$ are the last ticks at or before $v_i$
 and
$$\aligned \epsilon_{i,+}=X_{t_i}-X_{v_i}+Y_{s_i}-Y_{v_i}+\epsilon_i^{X}
+\epsilon_i^{Y},\\\epsilon_{i,-}=X_{t_i}-X_{v_i}-Y_{s_i}+Y_{v_i}+\epsilon_i^{X}
-\epsilon_i^{Y}.\endaligned $$

Note that
$$
  \widehat{\langle X,Y \rangle}_1 = \frac{1}{4}(\widehat{\langle
  Z^+,Z^+ \rangle}_1-\widehat{\langle Z^-,Z^- \rangle}_1).
$$
We can first prove analogues results as Theorem 1 for
$\widehat{\langle Z^+,Z^+ \rangle}_1$ and $\widehat{\langle Z^-,Z^-
\rangle}_1$, then utilize the results to obtain the final conclusion
for TSCV.

For $\widehat{\langle Z^+,Z^+ \rangle}_1$, the derivation is
different from that of Theorem 1 only for the terms that involve the
noise, namely $\sqrt{\bar{n}_K}R_1$ and $\sqrt{\bar{n}_K}R_2$. Write
$\tilde\Delta X_i=X_{t_i}-X_{v_i}$ and $\tilde\Delta
Y_i=Y_{s_i}-Y_{v_i}$. Then, we have, the first term in
$\sqrt{\bar{n}_K}R_1$ becomes
 $$\aligned &\frac{\sqrt{\bar{n}_{K}}\sqrt{\tilde n}}{K} \cdot
 \frac{2}{\sqrt{\tilde n}}\sum_{i=1}^{\tilde n }\epsilon_{i,+} \epsilon_{i-1,+}\\
 =& \frac{\sqrt{\bar{n}_{K}}\sqrt{\tilde n}}{K} \cdot
 \frac{2}{\sqrt{\tilde n}}\sum_{i=1}^{\tilde n} \Big(\tilde\Delta X_i\tilde\Delta X_{i-1}+\tilde\Delta X_i \tilde\Delta Y_{i-1}
 +\tilde\Delta X_i(\epsilon_{i-1}^X+\epsilon_{i-1}^Y) \\&+\tilde\Delta Y_i\tilde\Delta X_{i-1}
+\tilde\Delta Y_i \tilde\Delta Y_{i-1}
 +\tilde\Delta Y_i(\epsilon_{i-1}^X+\epsilon_{i-1}^Y)+(\epsilon_{i}^X+\epsilon_{i}^Y)\tilde\Delta X_{i-1} \\& +
 (\epsilon_{i}^X+\epsilon_{i}^Y)\tilde\Delta Y_{i-1}+
 (\epsilon_{i}^X+\epsilon_{i}^Y)(\epsilon_{i-1}^X+\epsilon_{i-1}^Y) \Big)\endaligned$$
The only $O_P(1)$ term is the last term, which involves only
independent normals, and can be dealt with by the same way as before
(again assume $\tilde n$ is even for the simplicity of presentation
below):
$$\aligned & E\exp \Big \{ 2\theta n^{-1/2} \sum_{\mbox{odd $i$}}
(\epsilon_{i}^X+\epsilon_{i}^Y)(\epsilon_{i-1}^X+\epsilon_{i-1}^Y)
\Big \}\\= & E\exp \Big \{ 2\theta {\tilde  n}^{-1/2} \sum_{\mbox{even
$i$}}
(\epsilon_{i}^X+\epsilon_{i}^Y)(\epsilon_{i-1}^X+\epsilon_{i-1}^Y)
\Big \} \\ = &
\left(\frac{1}{1-4(\eta_X^2+\eta_Y^2)^2\theta^2/\tilde n}\right)^{{\tilde n}/4}\\
 \leq &
\exp\{2(\eta_X^2+\eta_Y^2)^2\theta^2 \}, \mbox{ when } |\theta|\leq
\frac{\sqrt{\tilde  n}}{2\sqrt{2}(\eta_X^2+\eta_Y^2)}.
\endaligned$$
The other terms are of a smaller order of magnitude. By applying an
$a_{\tilde n}$ sequence which grows moderately with ${\tilde n}$ as
in the proof of Theorem 1 (we can set $a_{\tilde n}=\tilde
n^{1/12}$), we can see easily that their exact bounds don't have
effect on our choice of $C_{1}$, $C_2$ or $\omega$. All we need to
show is that the moment generating functions of these terms can
indeed be suitably bounded as (\ref{ineq:MGF}). To show this, first
note that, for any positive number  $a$ and real valued $b$, by the
optional sampling theorem (applied to sub-martingales $\exp(aB_s^2)$
and $\exp(b\tilde\Delta y B_s)$ with stopping time $[X]_u\leq
C_\sigma^2 u$ for real number $\tilde\Delta y$), we have,
\begin{equation}\label{ineq:tildeD} \aligned E \Big (\exp\{
a (\tilde \Delta X_{i})^2 \}|\mathcal{F}_{i-1}\Big )& \leq \Big (
E\big(\exp\{a C_\sigma^2C_\Delta Z^2 /{\tilde n}
\} \big) \Big)\mbox{ for } Z\sim N(0,1)\\
& =  \left(\frac{1}{1-2 a C_{\sigma}^2 C_{\Delta}/{\tilde
n}}\right)^{1/2},
\endaligned\end{equation}
where $\mathcal{F}_i$ is the information collected up to time $v_i$.
Inequality (\ref{ineq:tildeD}) holds when $\tilde \Delta X_{i}$ is
replaced by $\tilde \Delta Y_{i}$. Similarly,
\begin{equation}\label{ineq:tildeDD}
\aligned E \Big (\exp\{b \tilde \Delta X_{i} \tilde \Delta Y_{i-1}
\}|\mathcal{F}_{i-2}\Big )& \leq E \Big (E(\exp\{ b \tilde \Delta
X_{i} \tilde \Delta Y_{i-1}
\}|\mathcal{F}_{i-1})|\mathcal{F}_{i-2}\Big )\\
& \leq E \Big ( \exp\{ b^2 C_\Delta C_\sigma^2(\tilde\Delta
Y_{i-1})^2/2\tilde n \}|\mathcal{F}_{i-2}\Big )\\
& \leq  \left(\frac{1}{1- b^2 C_{\sigma}^4 C_{\Delta}^2/{\tilde
n^2}}\right)^{1/2} .
 \endaligned
\end{equation}
 The inequalities (\ref{ineq:tildeD}) and (\ref{ineq:tildeDD}) can be used to obtain the bounds we need. For example,
by (\ref{ineq:tildeDD}) and the law of iterated expectations,
 $$\aligned E\Big(\exp\{\theta
 \sum_{\mbox{odd i}} \tilde\Delta X_i\tilde\Delta
Y_{i-1}\}\Big) \leq & \left(\frac{1}{1- \theta^2 C_{\sigma}^4
C_{\Delta}^2/{\tilde n^2}}\right)^{\tilde{n}/4} \\ \leq &
\exp\left\{\theta^2 C_{\sigma}^4 C_{\Delta}^2/{2\tilde n}\right\}
\mbox{ when } |\theta|\leq \frac{\tilde n}{\sqrt{2}C_\sigma^2
C_\Delta};
\endaligned$$
by independence, normality of the noise, the law of iterated
expectations and (\ref{ineq:tildeD}), we have
$$
\aligned & E\Big(\exp\{\frac{ \theta }{a_{\tilde n}}
\sum_{i=1}^{\tilde n}\tilde\Delta
X_i(\epsilon_{i-1}^X+\epsilon_{i-1}^Y)\}\Big)\\
 =& E
\Big (\exp\{ \sum_{i=1}^{\tilde n}(\frac{\theta}{a_{\tilde n}}
\tilde \Delta X_{i})^2(\eta_X^2+\eta_Y^2)/2\}
\Big )\\
\leq &  \left(\frac{1}{1-(\eta_X^2+\eta_Y^2)\theta^2  C_{\sigma}^2
C_{\Delta}/{\tilde n a_{\tilde n}^2}}\right)^{\tilde n/2}\\
\leq & \exp\{(\eta_X^2+\eta_Y^2) C_{\sigma}^2
C_{\Delta}\theta^2/a_{\tilde n}^2\}, \mbox{ when } |\theta|\leq
\frac{\sqrt{{\tilde n}}a_{\tilde n}}{C_{\sigma}
\sqrt{2C_{\Delta}(\eta_X^2+\eta_Y^2)}}.
\endaligned
$$
Similar results can be found for the other terms above, with the
same techniques.

The second term in $\sqrt{\bar{n}_K}R_1$ works similarly and have
the same bound. The other terms in $\sqrt{\bar{n}_K}R_1$ and the
whole term of $\sqrt{\bar{n}_K}R_2$ are of order $o_P(1)$. Again, by
using a sequence $a_{\tilde n}$ we can conclude immediately that
their exact bounds won't matter in our choice of the constants and
we only need to show that their moment generating functions are
appropriately bounded as (\ref{ineq:MGF}). The arguments needed to
prove the inequalities of form (\ref{ineq:MGF}) for each elements in
these terms are similar to those presented in the above proofs, and
are omitted here.

Hence, by still letting $w=14$ and redefining
$$\aligned C_{1,x}&=\frac{1}{4(2C_\sigma)^2 C_\Delta} \;\;\;\mbox{ and }\\
 C_2&=\max\{2(2C_\sigma)^4C_\Delta^2, 2(\eta_X^2 +\eta_Y^2)^2\} \\
&= 32 C_\sigma^4 C_\Delta^2 \mbox{ for the typical case when }
C_\sigma\geq \eta_X, \eta_Y,\endaligned
$$
we have, when $0 \leq x \leq  c' {\tilde n}^{1/6}$,
$$\aligned
    P\{{\tilde n}^{1/6}|\widehat{\langle Z^+, Z^+ \rangle}_1-\int_0^1
    {\sigma_t^{Z^+}}^2 dt|>x\}
    \leq &  4 \exp(- C' x^2),\;\;\\
    \mbox{ and }  P\{{\tilde n}^{1/6}|\widehat{\langle Z^-, Z^- \rangle}_1-\int_0^1
    {\sigma_t^{Z^-}}^2 dt|>x\} \leq &  4 \exp(- C' x^2),\;
     \endaligned
$$ where  $$  c'=2C_{1,x}C_2 \mbox{ and } C'=(32C_2w^2)^{-1}.$$

Finally, for the TSCV estimator, when $0 \leq x \leq c {\tilde
n}^{1/6}$,
$$\aligned & P\{{\tilde{n}}^{1/6}|\widehat{\langle X,Y
\rangle}_1-\int_0^1\sigma_t^{(X)}\sigma_t^{(Y)}\rho_t^{(X,Y)}dt|>x\}\\
\leq & P\{{\tilde{n}}^{1/6}|\widehat{\langle Z^+, Z^+
\rangle}_1-\int_0^1
    {\sigma_t^{Z^+}}^2 dt|>2x\}\\ & +P\{{\tilde n}^{1/6}|\widehat{\langle Z^-, Z^- \rangle}_1-\int_0^1
    {\sigma_t^{Z^-}}^2 dt|>2x\}\\
    \leq &8 \exp(- C x^2),\;\;\;
     \endaligned
$$ where  \begin{equation} \label{MC_tscv} c=c'/2=C_{1,x}C_2 \mbox{ and }
C=4C'=(8C_2w^2)^{-1}.\end{equation} This completes the proof.

Note that the argument is not restricted to TSCV based on the
pairwise refresh times -- it works the same (only with $\tilde{n}$
replaced by $\tilde{n}_*$, the observation frequency of the
all-refresh method) for the case when the synchronization scheme is
chosen to be the all-refresh method, as long as the sampling
conditions Condition 3-4 are satisfied.

\end{document}